\documentclass[amssymb,aps,prd,floats,floatfix,nofootinbib,superscriptaddress,a4paper,10pt]{revtex4}
\usepackage[intlimits]{amsmath}
\usepackage{tensor}
\usepackage[dvipsnames]{xcolor}
\usepackage{subfigure,url}
\usepackage{mathtools}
\usepackage{hyperref}

\begin{document}

\title{A Monte Carlo approach to stationary kinetic disks in the Kerr spacetime}

\author{Ghafran Khan}
\email{ghafran.khan@doctoral.uj.edu.pl}
\affiliation{Szko{\l}a Doktorska Nauk \'{S}cis{\l}ych i Przyrodniczych, Uniwersytet Jagiello\'{n}ski}
\affiliation{Instytut Fizyki Teoretycznej, Uniwersytet Jagiello\'{n}ski, {\L}ojasiewicza 11, 30-348 Krak\'ow, Poland}
\author{Patryk Mach}
\email{patryk.mach@uj.edu.pl}
\affiliation{Instytut Fizyki Teoretycznej, Uniwersytet Jagiello\'{n}ski, {\L}ojasiewicza 11, 30-348 Krak\'ow, Poland}

\begin{abstract}
We extend a recently proposed Monte Carlo scheme for computing stationary solutions of the general-relativistic Vlasov equation to the Kerr spacetime. As an example, we focus on razor-thin configurations of a gas confined to the equatorial plane and extending to spatial infinity. We consider monoenergetic models as well as solutions corresponding to planar Maxwell-J\"{u}ttner distributions at infinity. In both cases, the components of the particle current surface density are recovered within the proposed Monte Carlo framework. Some aspects of razor-thin kinetic disk models, including an analysis of the bulk angular momentum and angular velocity, are briefly covered.
\end{abstract}

\maketitle

\section{Introduction}

The general-relativistic kinetic theory can be understood as an alternative to the more popular hydrodynamical or magnetohydrodynamical description, valid in the regime in which collisions between individual particles of the gas are rare and the assumption of the local themrodynamic equilibrium may not be satisfied. It was introduced in 1960's \cite{Tauber1961,Israel1963,Lindquist1966}, following seminal works of Synge \cite{Synge1957}. Soon after, Ehlers formulated the general-relativistic kinetic theory in a geometrical fashion, rooted in the geometry of the tangent bundle \cite{Ehlers1971a,Ehlers1971b,Ehlers1973}. Early applications of the general-relativistic kinetic theory to modeling of relativistic stellar systems can be found in \cite{ST1993,ST19932}. Among various astrophysical applications, one should mention models of dark matter \cite{DJA2017,pmao1,pmao2,pmao3,mms2025a} or particle-in-cell simulations of a collisionless magnetized plasma around black holes \cite{PPB2019, BRP2021, CCD2022, GPQ2023}. The kinetic description of the collisionless gas provides a well-known realistic matter model used in mathematical general relativity, where it is usually referred to as the Vlasov gas. Modern reviews can be found in \cite{OT2014,Acuna2022} (a general geometric approach), \cite{HA2011} (Einstein-Vlasov systems). Some aspects of the general-relativsitic kinetic theory are covered in \cite{CK2002}.

In this paper, we show that a Monte Carlo scheme proposed recently to compute stationary solutions of the general-relativistic Vlasov equation (collisionless Boltzmann equation) can be extended to the Kerr spacetime. The original method was introduced in \cite{MCO2023}, where it has been illustrated in the context of spherically symmetric (or axially symmetric and planar) models in the Schwarzschild spacetime. It was later generalized for planar models of accretion onto moving Schwarzschild black holes in \cite{CMO2024}.

As a convenient toy model in the Kerr spacetime, we consider razor-thin disks of a collisionless gas in the equatorial plane. They represent a kinetic, planar analog of the Bondi accretion model (steady, spherically symmetric accretion from infinity, where the gas is assumed to be homogeneous and at rest). In the planar case, the gas is assumed to extend to infinity in the equatorial plane.

General-relativistic Bondi-type accretion of the Vlasov gas onto black holes was extensively studied in spherical symmetry. The accretion rate of monoenergetic particles infalling spherically onto a Schwarzschild black hole was computed in \cite{Shapiro1983,Shapiro1985}. The first detailed analysis of the Bondi-type accretion of a Vlasov gas onto the Schwarzschild black hole was presented in \cite{Rioseco2017a,Rioseco2017b}. This model was later generalized for the Reissner-Nordstr\"{o}m spacetime in \cite{CM2020,Li2025} and for more exotic spacetimes in \cite{Liao2022,Cai2023,Zhang2025}. A generalization for a large class of spherically symmetric static spacetimes was shown in \cite{Momennia2025}. Spherically symmetric accretion onto a Schwarzschild black holes from a sphere of a finite radius was analyzed in \cite{Gamboa2021}.

Much less is known about analytic (or exact) kinetic models in the Kerr spacetime. Stationary kinetic disks confined to the equatorial plane were studied in \cite{CMO2022,Khan2025}. A dynamical scenario, leading to the so-called phase space mixing was investigated in \cite{Rioseco2018} for planar models and recently in \cite{Rioseco2024} without restrictions to the equatorial plane. Kinetic Bondi-type solutions in the Kerr spacetime were derived in \cite{Li2023,mms2025a,mms2025b} (see also \cite{Liu2025} for an analogous model in the Kerr-Newmann spacetime).

Our Monte Carlo procedure consists of the following three (essentially independent) elements: (i) solving geodesic equations describing the motion of individual particles of the gas (a part of the full solution is usually sufficient), (ii) selecting an appropriate sample of geodesic parameters, corresponding to the kinetic distribution function of the investigated model, (iii) introducing a suitable averaging scheme to provide macroscopic (observable) quantities. A novel aspect introduced in \cite{MCO2023} is related to this last element. We perform the averaging procedure by integrating a fine-grained distribution function corresponding to a set of individual particle trajectories over hypersurfaces adapted to symmetries of the problem. For stationary solutions, one can average over timelike hypersurfaces. Ultimately, our method is designed to work in cases in which controlling the phase-space ranges available for the motion of particles of the gas is difficult, and hence one has to resort to numerical methods. However, as with any numerical method, it is essential to provide a selection of reliable tests. In this work, we go beyond the assumptions of spherical symmetry of the spacetime, and show that the method can be generalized to axially symmetric spacetimes.

The order of this paper is as follows. In Section \ref{sec:vlasov} we review necessary elements of the general-relativistic kinetic theory. Section \ref{sec:geodesicmotion} discusses the geodesic motion confined to the equatorial plane of the Kerr spacetime in horizon-penetrating Kerr coordinates. In Section  \ref{sec:vlasovequatorial} we give the kinetic description of the gas in the equatorial plane of the Kerr spacetime. We recall basic elements of the planar accretion model considered in \cite{CMO2022}, introduce its monoenergetic variant, and discuss the properties of the bulk angular velocity. The Monte Carlo method is discussed in Sec.\ \ref{sec:montecarlo} and \ref{geodesic_parameters}. In Section \ref{sec:numresults} we provide a selection of numerical examples. Section \ref{sec:discussion} contains conclusions.

We work in standard geometric units with $c = G = 1$, where $c$ denotes the speed of light, and $G$ is the gravitational constant. The metric signature is $(-,+,+,+)$.

\section{Kinetic description}\label{sec:vlasov}

In this section we recall some basic notions regarding the general-reativistic kinetic description of gasses. For simplicity, we consider the so-called simple gas, consisting of identical spinless particles of rest mass $m > 0$. Let $(\mathcal M,g)$ denote the spacetime manifold endowed with the metric $g$. The gas is described in terms of the one-particle distribution function $\mathcal F \colon U \to \mathbb R$, where $U \subseteq T^\ast \mathcal M$. Here the cotangent bundle is defined as
\begin{equation}
    T^\ast \mathcal M = \bigcup_{x \in \mathcal M} \{ (x,p) \colon p \in T^\ast_x \mathcal M \},
\end{equation}
where $T^\ast_x \mathcal M$ denotes the cotangent space at $x \in \mathcal M$ (for a formulation based on the tangent bundle see, e.g., \cite{OT2014}). The distribution function $\mathcal F$ has the following statistical interpretation. Let $S$ be a spacelike hypersurface in $\mathcal M$. The number of particle trajectories whose projections on $\mathcal M$ intersect $S$ can be defined as
\begin{equation}
\mathcal N[S] = - \int_S \left[ \int_{P_x^+} \mathcal F(x,p) p_\mu s^\mu \mathrm{dvol}_x(p) \right] \eta_S,
\end{equation}
where $s$ is a future-directed unit vector normal to $S$, $\eta_S$ denotes the three-dimensional volume element induced on $S$. Here
\begin{equation}
    P_x^+ = \{ p \in T_x^\ast \mathcal M \colon g^{\mu \nu} p_\mu p_\nu < 0, \, p \text{ is future directed} \}
\end{equation}
and
 $\mathrm{dvol}_x(p)$ denotes the volume element in $P_x^+$,
\begin{equation}
\mathrm{dvol}_x(p) = \sqrt{- \mathrm{det} g^{\mu \nu}(x)} dp_0 dp_1 dp_2 dp_3.
\end{equation}
Defining the so-called particle current density
\begin{equation}
\mathcal J_\mu(x) = \int_{P_x^+} \mathcal F(x,p) p_\mu \mathrm{dvol}_x(p),
\end{equation}
one can write the expression for $\mathcal N[S]$ as
\begin{equation}\label{particlecurrentdensity}
\mathcal N[S] = - \int_S \mathcal J_\mu s^\mu \eta_S.
\end{equation}
In analogy to the particle current density $\mathcal J_\mu(x)$, the energy-momentum tensor of the Vlasov gas is defined as
\begin{equation}
\mathcal T_{\mu \nu} (x) = \int_{P_x^+} \mathcal F(x,p) p_\mu p_\nu \mathrm{dvol}_x(p). 
\end{equation}

If collisions between individual gas particles are neglected, the distribution function satisfies the following collisionless Boltzmann (or Vlasov) equation:
\begin{equation}
g^{\mu \nu} p_\nu \frac{\partial \mathcal F}{\partial x^{\mu}}
 - \frac{1}{2}\frac{\partial g^{\alpha\beta}}{\partial x^{\mu}}p_{\alpha}p_{\beta}
 \frac{\partial \mathcal F}{\partial p_{\mu}} = 0.
 \label{Eq:Vlasov}
\end{equation}
The characteristics of Eq.\ (\ref{Eq:Vlasov}) coincide with geodesic particle trajectories, satisfying geodesic equations, which, for convenience, we write as the Hamiltonian system
\begin{equation}
\label{eq:HamEqs}
    \frac{dx^\mu}{d\tau} = \frac{\partial \mathcal H}{\partial p_\mu}, \quad \frac{d p_\nu}{d \tau} = \frac{\partial \mathcal H}{\partial x^\nu}, 
\end{equation}
where the Hamiltonian $\mathcal H$ reads $\mathcal H (x,p) = \frac{1}{2} g^{\mu \nu}(x) p_\mu p_\nu$, and $\tau$ denotes an affine parameter. This is, of course, equivalent to saying that $\mathcal F$ remains constant along geodesics, which can be expressed as
\begin{eqnarray}
\label{vlasov}
    \frac{d \mathcal F}{d \tau}=\frac{\partial \mathcal F}{\partial x^\mu} \frac{d x^\mu}{d \tau}+\frac{\partial \mathcal F}{\partial p_\nu} \frac{d p_\nu}{d \tau}= \frac{\partial \mathcal F}{\partial x^\mu} \frac{\partial \mathcal H}{\partial p_\mu}-\frac{\partial \mathcal F}{\partial p_\nu} \frac{\partial \mathcal H}{\partial x^\nu} = \{\mathcal H, \mathcal F \} = 0,
\end{eqnarray}
where $\{._,.\}$ denotes the Poisson bracket. It is easy to check that Eq.\ (\ref{Eq:Vlasov}) or, equivalently, Eq.\ (\ref{vlasov}) implies the conservation laws $\nabla_\mu \mathcal J^\mu = 0$ and $\nabla_\mu \mathcal T^{\mu \nu} = 0$.

The form of Eq.\ (\ref{vlasov}) implies that any distribution function $\mathcal F$ that depends on $x^\mu$ and $p_\mu$ via integrals of the geodesic motion $I_1(x,p), \dots, I_s(x,p)$, i.e., $\mathcal F(x,p) = f(I_1(x,p), \dots, I_s(x,p))$ for some function $f$, would automatically satisfy Eq.\ (\ref{vlasov}). More complicated solutions can also be obtained by a suitable transformation to action-angle variables (see, e.g., \cite{pmao1,pmao2} for a practical example).

Equations (\ref{eq:HamEqs}) imply that $p^\mu = dx^\mu/d\tau$. In what follows we assume the normalization $\mathcal H(x,p) = \frac{1}{2}g^{\mu \nu}(x) p_\mu p_\nu = - \frac{1}{2} m^2$, where $m$ denotes the rest mass of a single particle. For $m > 0$ this implies that the parameter $\tau$ is related to the proper time $\tau_\mathrm{proper}$ by $\tau = \tau_\mathrm{proper}/m$.

\section{Geodesic motion in the Kerr Spacetime}
\label{sec:geodesicmotion}

Kerr geodesics play a double role in this work. On the one hand, a computation of geodesic trajectories of individual gas particles constitutes an important ingredient of the proposed Monte Carlo scheme. On the other hand, the geometry of the phase space associated with the geodesic motion is essential in the construction and understanding of the properties of kinetic models discussed in this paper.

\subsection{Separability}

In horizon-penetrating Kerr coordinates $(t,r,\vartheta,\varphi)$, the Kerr metric can be written as\footnote{Horizon penetrating Kerr coordinates are related to the more familiar Boyer-Lindquist ones $(t^\prime,r^\prime,\vartheta^\prime,\varphi^\prime)$ by
\begin{eqnarray}
    d t = dt^\prime +\frac{2Mr^\prime}{{r^\prime}^2 - 2 M r^\prime + a^2} dr^\prime,  \qquad  d  \varphi = d\varphi^\prime + \frac{a}{{r^\prime}^2 - 2 M r^\prime + a^2}dr^\prime.
\end{eqnarray}
In Boyer-Lindquist coordinates $(t^\prime,r^\prime,\vartheta^\prime,\varphi^\prime)$ the Kerr metric can be written as \cite{BoyerLindquist1967}
\begin{equation}
\label{kerr4d}
    g = -\left( 1 - \frac{2Mr^\prime}{\rho^2} \right) {dt^\prime}^2 - \frac{4 M a r^\prime \sin^2 \vartheta^\prime}{\rho^2} dt^\prime d\varphi^\prime + \frac{\rho^2}{\Delta} {dr^\prime}^2 + \rho^2 {d\vartheta^\prime}^2 + \left( {r^\prime}^2 + a^2 + \frac{2 M a^2 r^\prime \sin^2 \vartheta^\prime}{\rho^2} \right) \sin^2 \vartheta^\prime {d\varphi^\prime}^2.
\end{equation}}
\begin{eqnarray}
g & = & -d t^2 +dr^2 -2a\sin^2\vartheta dr d  \varphi 
 + \left( r^2 + a^2 \right)\sin^2 \vartheta d \varphi^2 
 + \rho^2 d\vartheta ^2 
 + \frac{2M r}{\rho^2}  \left(d t + dr - a\sin ^2 \vartheta d  \varphi \right)^2 \\
 & = & -\left(1-\frac{2Mr}{\rho^2}\right){dt}^2-\frac{4Mr}{\rho^2}a\sin^2\vartheta dt d\varphi + \frac{4Mr}{\rho^2} dt dr - 2 a \left(1 + \frac{2Mr}{\rho^2}\right)\sin^2\vartheta dr d\varphi
    \nonumber\\
    & & + \left(1+\frac{2Mr}{\rho^2}\right) dr^2 + \rho^2 d\vartheta^2 + \frac{\left[ (r^2+a^2)^2-a^2 \Delta \sin^2\vartheta \right]\sin^2\vartheta}{\rho^2} {d\varphi}^2.
\label{Eq:Kerr}
\end{eqnarray}
where
\begin{equation}
\rho^2 = {r}^2 + a^2 \cos^2 \vartheta, \qquad \Delta = r^2 - 2 M r + a^2.
\end{equation}
Here, $M$ denotes the black hole mass, and $a$ is the black hole spin parameter. We assume that $- M  < a < M$, i.e., we only consider the subextremal family of Kerr solutions. The inverse metric has the form
\begin{equation}
(g^{\mu\nu}) = \frac{1}{\rho^2}\left( \begin{array}{cccc}
 -(\rho^2 + 2M r) & 2Mr & 0 & 0 \\
 2Mr & \Delta & 0 & a \\
 0 & 0 & 1 & 0 \\
 0 & a & 0 & \sin^{-2}\vartheta
\end{array} \right),
\label{Eq:KerrInverse}
\end{equation}
and its determinant reads
\begin{equation}
\mathrm{det} \, g^{\mu \nu} = - \frac{1}{\rho^4 \sin^2 \vartheta}.
\end{equation}
The two zeros of $\Delta$, $r_\pm = M \pm \sqrt{M^2 - a^2}$, correspond to the radii of the inner (Cauchy) and the outer (event) horizon. The Kerr metric admits the following two Killing vectors, which in the horizon-penetrating Kerr coordinates can be expressed as
\begin{equation}
\label{killing}
k = \partial_t, \qquad \chi = \partial_\varphi.
\end{equation}
Timelike geodesic can be described  by Hamilton's equations (\ref{eq:HamEqs}) with the Hamiltonian 
\begin{equation}
    \mathcal{H}=\frac{1}{2}g^{\mu\nu}(x^\alpha)p_\mu p_\nu=\frac{1}{2\rho^2}\Big[-(\rho^2+2Mr)p_t^2+4Mrp_tp_r +\Delta p_r^2+p_\vartheta^2+2ap_rp_\varphi +\frac{1}{\text{sin}^2 \vartheta}p_\varphi^2\Big].
\end{equation}
In this case, Eqs.\ (\ref{eq:HamEqs}) posses the following constants of motion: $m^2 = - g^{\mu \nu} p_\mu p_\nu$, $E = - p_t$, $l_z = p_\varphi$, and
\begin{equation}
\label{carterconst}
    l^2 = p_\vartheta^2 + \left( \frac{p_\varphi}{\sin \vartheta} + a \sin\vartheta  p_t \right)^2 + m^2 a^2 \cos^2 \vartheta,
\end{equation}
and they can be separated as follows
\begin{subequations}
\begin{eqnarray}
    p_t & = & -E, \\
    (\Delta p_r - 2 M E r + a l_z)^2 & = & R(r), \\
    p_\vartheta^2 & = & l^2 - \left( \frac{p_\varphi}{\sin \vartheta} + a \sin \vartheta  p_t \right)^2 - m^2 a^2 \cos^2 \vartheta, \label{pvartheta} \\
    p_\varphi & = & l_z,
\end{eqnarray}
\end{subequations}
where
\begin{equation}
R(r) = \left[ E (r^2 + a^2) - a l_z \right]^2 - \Delta (l^2 + m^2 r^2).
\end{equation}
For practical reasons, in Eq.\ (\ref{carterconst}) the Carter constant \cite{Carter1968} is replaced by the constant $l$, which in the limit of $a \to 0$ reduces to the total angular momentum.

In the remainder of this paper we will use the following dimensionless quantities:
\begin{equation} \label{dimless}
r = M \xi, \quad a = M \alpha,\quad t =M T, \quad  \quad E = m \varepsilon, \quad l = M m \lambda, \quad\quad l_z = M m \lambda_z.
\end{equation}
Using the above definitions, one has
\begin{equation}
    R(r)=M^4 m^2 \tilde{R}, \qquad \Delta=M^2 \tilde{\Delta},
\end{equation}
where
\begin{equation}
    \tilde R = \left[ \varepsilon (\xi^2 + \alpha^2) - \alpha \lambda_z\right]^2 - \tilde \Delta (\lambda^2 + \xi^2), \qquad \tilde \Delta = \xi^2 - 2 \xi + \alpha^2.
\end{equation}
In the following, we will also denote $\xi_\pm = r_\pm/M = 1 \pm \sqrt{1 - \alpha^2}$, etc.

\subsection{Geodesic motion at the equatorial plane}

Restricting ourselves to the motion in the equatorial plane, we set $p_\vartheta = 0$, $\vartheta = \pi/2$. Equation (\ref{pvartheta}) then gives
\begin{equation}
l^2 = (p_\varphi + a p_t)^2 = (l_z - a E)^2.
\end{equation}
In what follows, we will adopt a convention with $l \ge 0$ and
\begin{equation}
\label{epssigmaconvention}
l_z = \epsilon_\sigma l + aE,
\end{equation}
where $\epsilon_\sigma = \pm 1$ denotes a sign.

In this case, the relevant covariant momentum components can be expressed as
\begin{subequations}
\label{momentumplanar}
\begin{eqnarray}
    p_t & = & - E, \\
    p_r & = & \frac{2 M r E - a l_z + \epsilon_r \sqrt{R(\xi)}}{\Delta} = \frac{2 M r E - \epsilon_\sigma a l - a^2 E + \epsilon_r \sqrt{R(\xi)}}{\Delta}, \\
    p_\varphi & = & \epsilon_\sigma l + e E,
\end{eqnarray}
\end{subequations}
where $\epsilon_r = \pm 1$ corresponds to the direction of the radial motion, and
\begin{equation}
    R(r) = ( E r^2  - a \epsilon_\sigma l  )^2 - \Delta (l^2 + m^2 r^2).
\end{equation}
Expressing Eqs.\ (\ref{momentumplanar}) in terms of dimensionless quantities, one obtains
\begin{subequations}
\begin{eqnarray}
    p_t & = & - m \varepsilon, \\
    p_r & = & m \frac{2 \xi \varepsilon - \epsilon_\sigma \alpha \lambda - \alpha^2 \varepsilon + \epsilon_r \sqrt{\tilde R(\xi)}}{\tilde \Delta}, \\
    p_\varphi & = & M m (\epsilon_\sigma \lambda + \alpha \varepsilon),
\end{eqnarray}
\end{subequations}
where
\begin{equation}
    \tilde{R} = (\xi^2 \varepsilon-\alpha \epsilon_{\sigma}\lambda)^2-(\xi^2-2\xi+\alpha^2)( \xi^2+\lambda^2).
\end{equation}

For the orbit restricted to the equatorial plane the geodesic equations can be written as
\begin{subequations}
\label{eomequator}
\begin{eqnarray}
\frac{dt}{d\tau} & = & p^t = \frac{(r + 2M) E + 2 M p_r}{r}, \\
\frac{dr}{d \tau} & = & p^r = \frac{\Delta p_r - 2 M E r + a l_z}{r^2}  = \epsilon_r \frac{\sqrt{R(r)}}{r^2}, \\
\frac{d\varphi}{d \tau} & = & p^\varphi = \frac{l_z + a p_r}{r^2}.
\end{eqnarray}
\end{subequations}
The geodesic motion confined to the equatorial plane can also be analyzed using the $(2+1)$-dimensional metric $\gamma_{\mu\nu}$ induced in the equatorial plane. It reads (in horizon-penetrating Kerr coordinates)
\begin{eqnarray}
    \gamma=-dt^2+dr^2-2adrd\varphi+(r^2+a^2)d\varphi^2+\frac{2M}{r}(dt+dr-ad\varphi)^2.
\end{eqnarray}
The inverse metric can be written as 

\begin{equation}
(\gamma^{\mu\nu}) = \frac{1}{r^2}\left( \begin{array}{cccc}
 -(r^2 + 2M r) & 2Mr  & 0 \\
 2Mr & \Delta  & a \\
 0 & a &  1
\end{array} \right),
\label{Eq: KerrInverse eq plane}
\end{equation}
where the coordinates are ordered as $(t,r,\varphi)$. It is easy to check that $\sqrt{-\mathrm{det} \, \gamma^{\mu\nu}}=1/r.$

A convenient step in solving Eqs.\ (\ref{eomequator}) consists of introducing the so-called Mino time $\tilde s$ \cite{Mino2003}, defined by
\begin{eqnarray}\label{minotime123}
    \tau = \int^{\tilde{s}}_0 r^{2} ds, 
\end{eqnarray}
or equivalently, $r^2 d x^\mu/d \tau = dx^\mu/d \tilde s$. Reparametrizing the geodesics with the Mino time, we get
\begin{subequations}
\label{eqs:geodesic2}
    \begin{eqnarray}
    \frac{d t}{d \tilde{s}}&=& \frac{\left(r^2+a^2\right)\left(r^2E-a\epsilon_\sigma l\right)}{\Delta}+a \epsilon_\sigma l + \epsilon_r\frac{2 M r}{\Delta} \sqrt{R(r)}, \label{time2} \\
     \frac{d r}{d \tilde{s}}&=& \epsilon_r \sqrt{R(r)}, \label{radial2} \\
      \frac{d \varphi}{d \tilde{s}}&=& \frac{a\left(r^2E-a\epsilon_\sigma l\right)}{\Delta}+\epsilon_\sigma l + \epsilon_r \frac{a}{\Delta} \sqrt{R(r)}. \label{curlyphi2} 
\end{eqnarray}
\end{subequations}
  It is convenient to work in the dimensionless variables (\ref{dimless}).  Accordingly, Eqs.\ (\ref{eqs:geodesic2}) can be written as
\begin{subequations}\label{geodesicminonormalize}
    \begin{eqnarray}
    \frac{d T}{d s}&=& \frac{\left(\xi^2+\alpha^2\right)\left(\xi^2\varepsilon-\alpha\epsilon_\sigma \lambda\right)}{\xi^2-2\xi+\alpha^2}+\alpha \epsilon_\sigma \lambda + \epsilon_r \frac{2 \xi}{\xi^2-2\xi+\alpha^2} \sqrt{\tilde{R}}, \\
     \frac{d \xi}{d s}&=& \epsilon_r \sqrt{\tilde{R}}, \label{dimradial} \\
     \frac{d \varphi}{d s}&=& \frac{\alpha\left(\xi^2\varepsilon-\alpha\epsilon_\sigma \lambda\right)}{\xi^2-2\xi+\alpha^2}+\epsilon_\sigma \lambda + \epsilon_r \frac{\alpha}{\xi^2-2\xi+\alpha^2} \sqrt{\tilde{R}}, \label{Azimuthslmotion}
\end{eqnarray}
\end{subequations}
where we have introduced the dimensionless Mino time $s = M m \tilde s$. Solutions of Eqs.\ (\ref{geodesicminonormalize}) were recently obtained in \cite{BLUCM2025} in terms of Weierstrass functions. In the following, we will only need explicit expressions for $\xi(s)$ and $\varphi(s)$.

To solve the radial equation (\ref{dimradial}), it is convenient to note that $\tilde R$ is a fourth-order polynomial with respect to $\xi$:
\begin{equation}
    \tilde R(\xi) = a_0 \xi^4 + 4 a_1 \xi^3 + 6 a_2 \xi^2 + 4 a_3 \xi + a_4,
\end{equation}
where
\begin{equation}
a_0 = \varepsilon^2 - 1,\quad a_1=\frac{1}{2},\quad
a_2= -\frac{1}{6}\left(\alpha^2+2 \epsilon_{\sigma} \alpha \varepsilon \lambda+\lambda^2\right),\quad
a_3=\frac{1}{2} \lambda^2, \quad a_4=0.
\end{equation}
Weierstrass invariants of $\tilde R(\xi)$ are given by
\begin{equation}
\label{WeierstrassInvariants}
g_2 = 3a_2^2-4a_1a_3, \quad g_3=2a_1a_2a_3-a_0a_3^2-a_2^3,
\end{equation}
where we have used the fact that $a_4 = 0$. A general solution of Eq.\ (\ref{dimradial}) can be written in the form \cite{CM2022a,CM2023,AEP2023,BLUCM2025}
\begin{eqnarray}\label{WeierstrassFormula}
    \xi(s)=\xi_{0}+\frac{-\epsilon_{r,0} \sqrt{\tilde{R}(\xi_0)}\wp^\prime_{\tilde{R}}(s)+\frac{1}{2}\tilde{R}^\prime(\xi_0)[\wp_{\tilde{R}}(s)-\frac{1}{24}\tilde{R}^{\prime\prime}(\xi_{0})]+\frac{1}{24}\tilde{R}(\xi_0)\tilde{R}^{\prime\prime\prime}(\xi_0)}{2[\wp_{\tilde{R}}(s)-\frac{1}{24}\tilde{R}^{\prime\prime}(\xi_0)]^2-\frac{1}{48}\tilde{R}(\xi_0)\tilde{R}^{(4)}(\xi_0)},
\end{eqnarray}
where $\wp_{\tilde{R}}(s)=\wp(s;g_{2},g_{3})$ and $\wp$ denotes the Weierstrass function. Here $\xi_0$ refers to the initial radius, i.e., $\xi(0) = \xi_0$. The sign $\epsilon_{r,0}$ corresponds to the radial direction of motion at $s = 0$. Note that Eq.\ (\ref{WeierstrassFormula}) represents an exact solution, regardless of the type of the orbit (existence, number, and locations of radial turning points, etc.).

Given the solution for $\xi(s)$, the azimuthal equation (\ref{Azimuthslmotion}) can be solved by quadratures as
\begin{equation}
\varphi(s)-\varphi(0) = \alpha \int^s_0 \frac{ \left(\varepsilon\xi^2(\bar{s})-\alpha \epsilon_\sigma \lambda\right)}{\xi^2(\bar{s})-2\xi(\bar{s})+\alpha^2} d\bar{s}+\alpha\int^s_0 \frac{ d \xi(\bar{s})/d\bar{s}} {\xi^2(\bar{s})-2\xi(\bar{s})+\alpha^2} d\bar{s}+ \epsilon_\sigma \lambda s.
\label{phiquadraturesb}
\end{equation}
This can be done analytically (see, e.g., \cite{AEP2023} for a calculation in Boyer-Lindquist coordinates), but the resulting expressions are lengthy. On the other hand, integrals appearing in Eq.\ (\ref{phiquadraturesb}) can be easily evaluated numerically, as in \cite{BLUCM2025}. An expression equivalent to Eq.\ (\ref{phiquadraturesb}), but explicitly regular at the black hole horizon for ingoing trajectories, reads \cite{BLUCM2025}
\begin{equation}
    \varphi(s) - \varphi(0) = \alpha \int_0^s \frac{\xi^2(\bar s) + \lambda^2}{\varepsilon \xi^2(\bar s) - \alpha \epsilon_\sigma \lambda - d \xi(\bar s)/d \bar s} d \bar s + \epsilon_\sigma \lambda s.
\label{phiquadraturesc}
\end{equation}
Note that the derivative $d \xi(\bar s)/d \bar s$ in Eqs.\ (\ref{phiquadraturesb}) and (\ref{phiquadraturesc}) reads $d \xi(\bar s)/d \bar s = \epsilon_r \sqrt{\tilde R(\xi(\bar s))}$. However, it is more convenient to obtain $d \xi(\bar s)/d \bar s$ by directly differentiating Eq.\ (\ref{WeierstrassFormula}), as this allows one to avoid the calculation of radial turning points (establishing the sign $\epsilon_r$).

Since, in particular examples, the Mino time can be hard to control, it is also convenient to invert the relation (\ref{WeierstrassFormula}). In this case, explicitly real formulas depend on the radial type of the trajectory. In what follows, we only consider unbound trajectories with $\varepsilon > 1$.  They can either plunge into the black hole (we refer to such trajectories as absorbed ones) or, if the angular momentum is large enough, they can be scattered by the centrifugal potential (these trajectories will be referred to as scattered ones). The distinction between absorbed and scattered trajectories is based on the properties of the radial potential $\tilde R(\xi)$. There are no (real) zeros of $\tilde R(\xi)$ outside the black hole horizon (i.e., for $\xi > \xi_+$) for absorbed trajectories, and there exists at least one zero of $\tilde R(\xi)$ for $\xi > \xi_+$ for scattered ones. Note, however, that for equatorial timelike geodesics, $\xi = 0$ is always a zero of $\tilde R(\xi)$. This also happens for $\alpha = 0$ (Schwarzschild case), but it is not in general true for generic Kerr geodesics. There exist timelike geodesics in the Kerr spacetime (not confined to the equatorial plane) for which $\tilde R(\xi)$ does not possess any real zeros.

Equation (\ref{WeierstrassFormula}) can be inverted by first considering the integral
\begin{eqnarray}\label{fourthorderploy}
S(\xi_0,\xi_1)=\int^{\xi_1}_{\xi_{0}} \frac{d\xi}{\sqrt{\tilde{R}(\xi)}},     
\end{eqnarray}
where $\tilde R(\xi) \ge 0$ for $\xi_0 \le \xi \le \xi_1$. A calculation of this integral can be found in \cite{CM2022a}, for the case with $\alpha = 0$, but it can also be adapted to equatorial Kerr geodesics.

For geodesics in the equatorial plane, $\tilde R(\xi)$ can be factored as $\tilde R(\xi) = \xi \psi(\xi)$, where $\psi(\xi)$ is a cubic polynomial (we adapt the notation from \cite{ONEILL1995}). Assume that $\varepsilon > 1$ and $\lambda \neq 0$. Then $\psi(\xi) \to - \infty$ for $\xi \to -\infty$, $\psi(\xi) \to + \infty$ for $\xi \to +\infty$, and $\psi(0) = 2 \lambda^2 > 0$. Thus, the polynomial $\psi(\xi)$ has a real zero in $(-\infty,0)$. The remaining zeros of $\psi(\xi)$ can be both real, or both complex (for simplicity, we exclude special cases with zeros of multiplicity higher than one), depending on the sign of its discriminant. In principle, if all zeros of $\psi(\xi)$ are real, they can either be all negative, or two of them must be positive. The latter situation happens necessarily for scattered orbits, but it is also possible for absorbed trajectories, if two real zeros of $\psi(\xi)$ are located in the interval $(0,\xi_-)$. It is possible to show that $\psi(\xi)$ cannot have three distinct negative roots. Thus, for scattered unbound orbits, $\tilde R(\xi)$ has 4 real zeros (one negative, $\xi = 0$, and two positive). For absorbed trajectories, $\tilde R$ can either have 4 real zeros (one negative, $\xi = 0$, and two positive satisfying $\xi < \xi_-$) or two real zeros (one negative and $\xi = 0$) and two complex ones.

Consider a scattered orbit. Let $\xi_2$ be the largest of the zeros of $\tilde R(\xi)$ (the radial turning point of the unbound scattered trajectory). For $\xi_2 \le \xi_0$, the integral $S(\xi_0,+\infty)$ can be expressed as
\begin{eqnarray}
\label{S1}
S_1(\xi_0,+\infty)=\frac{1}{\sqrt{y_3-y_1}}\left[\tilde{F}\left(\mathrm{\arccos}\sqrt{\frac{y_2-\left(\frac{a_2}{2}+\frac{a_3}{\xi_{0}}\right)}{y_2-y_1}} ,k \right)-\tilde{F}\left(\mathrm{\arccos}\sqrt{\frac{y_2-\frac{a_2}{2}}{y_2-y_1}},k \right)\right],
\end{eqnarray}
where, $y_1,y_2,$ and $y_3$ denote the roots of the polynomial $4y^3-g_2y-g_3,$ satisfying $y_1<y_2<y_3$, $k = \sqrt{(y_2-y_1)/(y_3-y_1)}$, and the Weierstrass invariants $g_2$, $g_3$ are given by Eq.\ (\ref{WeierstrassInvariants}). Here, $\tilde F(\phi,k)$ denotes the Legendre elliptic integral defined by
\begin{eqnarray}
\tilde F(\phi,k)=\int^\phi_0 \frac{d\chi}{\sqrt{1-k^2\text{sin}^{2}\chi}}, \qquad -\frac{\pi}{2}<\phi<\frac{\pi}{2}.
\end{eqnarray}
For general $\xi_0$ and $\xi_1$ satisfying $\xi_2 \le \xi_0 \le \xi_1$, the integral $S(\xi_0,\xi_1)$ can be computed as $S(\xi_0,\xi_1) = S_1(\xi_0,+\infty) - S_1(\xi_1,+\infty)$. For a motion from $\xi_0$ to $\xi_1$ for which $\epsilon_r$ remains constant, one then obtains $s(\xi_1) - s(\xi_0) = \epsilon_r S(\xi_0,\xi_1)$. The passage of the Mino time corresponding to the motion along a scattered trajectory from $\xi_0$ to $\xi_2$ (the radial turning point) with $\epsilon_r = -1$ and back from $\xi_2$ to $\xi_0$ with $\epsilon_r = +1$, can be expressed as $2 S(\xi_2,\xi_0)$.

The same formulas also work for an absorbed trajectory, provided that $\tilde R(\xi)$ admits four real zeros. If only two real zeros of $\tilde R(\xi)$ are present, the analogous formula reads 
\begin{eqnarray}
\label{S2}
    S_2(\xi_0,+\infty) & = & \frac{1}{2\sqrt{\mu}}\left[\tilde{F}\left(2\mathrm{arctan}\sqrt{\frac{\frac{a_2}{2}+\frac{a_3}{\xi_0}-y_1}{\mu}}, \tilde k \right) -\tilde{F} \left(2 \mathrm{\arctan}  \sqrt{\frac{\frac{a_2}{2}-y_1}{\mu}},\tilde k \right)  \right],
\end{eqnarray}
Here, $y_1$ denotes the real zero of the polynomial $4y^3-g_2y-g_3=4(y-y_1)(y^2+py+q).$ Therefore, $p^2-4q<0,$ and thus $y^2+py+q>0.$ The expressions for $\mu$ and $\tilde k$ are given as
\begin{eqnarray}
    \mu=\sqrt{y_1^2+py_1+q}, \quad \tilde k^2 = \frac{1}{2}\left( 1-\frac{y_1+p/2}{\mu} \right).
\end{eqnarray}

\section{Kinetic gas in the equatorial plane}
\label{sec:vlasovequatorial}

\subsection{Surface densities}

For the kinetic description of the gas confined to the equatorial plane, it is convenient to introduce the following quantities:
\begin{equation}
\label{surfacedens}
\mathcal F(t,r,\vartheta,\varphi;p_\alpha) = \delta(z) F(t,r,\varphi;p_\alpha), \quad \mathcal J_\mu (t,r,\vartheta,\varphi) = \delta(z) J_\mu(t,r,\varphi), \quad \mathcal T_{\mu \nu} (t,r,\vartheta,\varphi) = \delta(z) T_{\mu \nu} (t,r,\varphi),
\end{equation}
where the $z$ coordinate is chosen in such a way that the vector $\partial_z$ is of unit length and is normal to the equatorial plane $\vartheta = \pi/2$. Note that the coordinate $z$ must only be defined in the vicinity of the equatorial plane. Consequently, it suffices to take the relation $z = r \cos \vartheta$. This gives $\delta(z) = \delta(\vartheta-\pi/2)/r$. With these definitions,
\begin{equation}
\label{jmutmunusurface}
J_\mu(t,r,\varphi) = \int_{P_x^+} F(t,r,\varphi;p_\alpha) p_\mu \mathrm{dvol}_x(p), \quad T_{\mu \nu} (t,r,\varphi) = \int_{P_x^+} F(t,r,\varphi;p_\alpha) p_\mu p_\nu \mathrm{dvol}_x(p).
\end{equation}

The consistency condition for the motion confined to the equatorial plane is that
\begin{equation}
F(t,r,\varphi,p_t,p_r,p_\vartheta,p_\varphi) = \delta(p_z) f(t,r,\varphi,p_t,p_r,p_\varphi), 
\end{equation}
where $p_z$ is the momentum component associated with the coordinate $z$. It ensures that the particles have no momentum perpendicular to the equatorial plane, and consequently remain at $\vartheta = \pi/2$. Since at the equatorial plane $p_z = - p_\vartheta/r$, one has $\delta(p_z) = r \delta(p_\vartheta)$. Note that at the equatorial plane the momentum space volume element reduces to
\begin{equation}
\mathrm{dvol}_x(p) = \frac{1}{\rho^2 \sin \vartheta} dp_t dp_r dp_\vartheta dp_\varphi = \frac{1}{r^2}  dp_t dp_r dp_\vartheta dp_\varphi.
\end{equation}
This gives
\begin{equation}
\label{jmutmunusurface2}
J_\mu(t,r,\varphi) = \frac{1}{r} \int_{\bar P_x^+} f(t,r,\varphi, p_t, p_r, p_\varphi) p_\mu dp_t dp_r dp_\varphi, \quad T_{\mu \nu} (t,r,\varphi) = \frac{1}{r} \int_{\bar P_x^+} f(t,r,\varphi, p_t,p_r,p_\varphi) p_\mu p_\nu dp_t dp_r dp_\varphi,
\end{equation}
where
\begin{equation}
    \bar P_x^+ = \{ (p_t,p_r,p_\varphi) \colon \gamma^{\mu \nu} p_\mu p_\nu < 0, \, p \text{ is future-directed} \}.
\end{equation}

In the remainder of this paper, we will also use the covariant surface density, defined as
\begin{equation}
\label{ns}
    n_s = \sqrt{- \gamma_{\mu \nu} J^\mu J^\nu}.
\end{equation}

\subsection{Volume element in the momentum space}

In what follows, it is convenient to express the momentum volume element in terms of constants of motion. This can be done using the following formulas:
\begin{subequations}
\begin{eqnarray}
    m^2 & = & p^2_t-p^2_r - \frac{(a p_r+p_\varphi)^2}{r^2}+\frac{2M(p_r-p_t)^2}{r}, \\
    E & = & - p_{t}, \label{ptheta} \\
    l & = & \epsilon_{\sigma} (p_{\varphi}+ap_{t}).
\end{eqnarray}
\end{subequations}
The Jacobian determinant is defined by
\begin{equation}
J=\frac{\partial (m^2,E,l)}{\partial (p_t,p_r,p_\varphi)} = - \frac{2 \epsilon_{\sigma} (\Delta p_r + 2 M r p_t + a p_\varphi)}{r^2} = - \frac{2 \epsilon_{\sigma} \epsilon_r \sqrt{R(r)}}{r^2}.
\end{equation}
Therefore, the volume element appearing in Eqs.\ (\ref{jmutmunusurface2}) can be expressed as
\begin{equation}
    dp_t dp_r dp_\varphi = \frac{r^2 m d m dE d l}{\sqrt{R(r)}}.
\end{equation}
Equivalently,
\begin{equation}
\sqrt{- \mathrm{det} \, \gamma^{\mu \nu}} dp_t dp_r dp_\varphi =  \frac{r}{\sqrt{R(r)}} m dm dE ldl,
\end{equation}
or, in terms of dimensionless quantities,
\begin{equation}
\sqrt{- \mathrm{det} \, \gamma^{\mu \nu}} dp_t dp_r dp_\varphi =  \frac{ \xi m^2}{\sqrt{\tilde R(\xi)}} dm d\varepsilon d\lambda.
\end{equation}

\section{Accretion Model}

As a convenient toy model allowing to test the proposed Monte Carlo scheme outside spherical symmetry, we consider a planar accretion model discussed in \cite{CMO2022}.

Within the equatorial plane, the gas is assumed to extend to infinity, where it remains at rest and is homogeneous. We restrict ourselves to the so-called simple gas, consisting of same-mass particles, and consider two cases with distribution functions depending on the energy only: monoenergetic and Maxwell-J\"{u}ttner models. For simplicity, we only take into account unbound particle trajectories, both absorbed and scattered ones.

Unbound absorbed trajectories can be found for any $\varepsilon > 1$. They are characterized by the condition $\lambda < \lambda_c(\varepsilon,\alpha,\epsilon_\sigma)$, where $\lambda_c(\varepsilon,\alpha,\epsilon_\sigma)$ is expressed in a parametric form as
\begin{equation}
    \lambda_c = \frac{\xi^{5/4} - \alpha \epsilon_\sigma \xi^{3/4}}{\sqrt{2 \alpha \epsilon_\sigma + (\xi - 3)\sqrt{\xi}}}, \qquad \varepsilon = \frac{\alpha \epsilon_\sigma + (\xi - 2) \sqrt{\xi}}{\xi^{3/4}\sqrt{2 \alpha \epsilon_\sigma + (\xi - 3) \sqrt{\xi}}}.
\end{equation}
A scattered unbound trajectory can reach a given radius $\xi$, provided that there exists a radial turning point (a zero of $\tilde R$) in the range $(\xi_+,\xi)$. This amounts to a condition $\varepsilon \ge \varepsilon_\mathrm{min}(\xi,\alpha,\epsilon_\sigma)$, where
\begin{equation}
\label{eminlimit}
    \varepsilon_\mathrm{min}(\xi,\alpha,\epsilon_\sigma) = 
\begin{cases}
    \infty &   \xi<\xi_{\mathrm{ph}},  \\
    \sqrt{\frac{-2\epsilon_\sigma\alpha\left(\alpha^2+(\xi-2)\xi\right)\xi^{-1/2}+\alpha^2(5-3\xi)+(\xi-3)(\xi-2)^2\xi}{\xi\left((\xi-3)^2\xi-4\alpha^2\right)}}  & \xi_{\mathrm{ph}}<\xi<\xi_{\mathrm{mb}},  \\
    1 \quad  & \xi\ge\xi_{\mathrm{mb}}.
\end{cases}
\end{equation}
Here $\xi_\mathrm{ph}$ and $\xi_\mathrm{mb}$ denote, respectively, the dimensionless radii of the circular photon orbit and the marginally bound orbit. They are given by $\xi_{\mathrm{ph}}=2+2\cos\left[2\arccos(-\epsilon_\sigma \alpha)/3\right]$ and $\xi_\mathrm{mb} = 2 - \epsilon_\sigma \alpha + 2 \sqrt{1 - \epsilon_\sigma \alpha}$. For unbound scattered orbits, the parameter $\lambda$ ranges from $\lambda_c(\varepsilon,\alpha,\epsilon_
\sigma)$ to $\lambda_\mathrm{max}(\xi,\varepsilon,\epsilon_\sigma)$, where
\begin{equation}
\label{lambdamax} 
    \lambda_\mathrm{max}  (\xi,\varepsilon,\epsilon_\sigma) = \frac{\xi}{\xi-2}  \Bigg \{ \sqrt{\alpha^2 \left(\varepsilon^2 + \frac{2}{\xi}-1 \right) +(\xi-2)\left[ \xi(\varepsilon^2-1) + 2 \right] }-\epsilon_\sigma \alpha \varepsilon \Bigg \}.
\end{equation}

\subsection{Monoenergetic distributions}

The monoenergetic distribution function $f$ at the equatorial plane can be written as
\begin{equation}
    f=A\delta(m-m_0)\delta(\varepsilon-\varepsilon_0),
\end{equation}
in the phase-space region occupied by absorbed and scattered trajectories described in the previous section. We assume that $f \equiv 0$ on the complementary region of the phase space (which includes, e.g., bound orbits or orbits emanating from the white hole). Here $m_0$ is the particle mass and $A$ denotes the normalization constant (with the dimension of $M^{-2} m_0^{-3}$). While monoenergetic flows were not discussed in \cite{CMO2022}, they serve as a convenient illustration of various dynamical aspects of the model. Monoenergetic kinetic flows in the Kerr spacetime were also discussed in \cite{Khan2025} and \cite{mms2025b}.

The components of the particle current surface density $J_\mu$ can be expressed as $J_\mu = J_\mu^\mathrm{(abs)} + J_\mu^\mathrm{(scat)}$, where the components $J_\mu^\mathrm{(abs)}$ and $J_\mu^\mathrm{(scat)}$ consist of absorbed and scattered trajectories, respectively. According to the characterization of the phase-space ranges discussed in \cite{CMO2022} and summarized above, they can be computed as follows:
\begin{subequations}\label{monoenergeticanalyticresult}
\begin{eqnarray}
    J_{t}^{\mathrm{(abs)}}(\xi)&=& - A m_{0}^{3} \xi \sum_{\epsilon_{\sigma}=\pm1} \int^{\infty}_{1} d\varepsilon \delta(\varepsilon-\varepsilon_{0})\varepsilon \int^{\lambda_{c}(\varepsilon,\alpha,\epsilon_\sigma)}_0 \frac{d\lambda}{\sqrt{\tilde{R}(\xi)}} = - A  m^3_{0} \xi \varepsilon_{0} \Theta(\varepsilon_{0}-1) \sum_{\epsilon_{\sigma}=\pm1}  \int^{\lambda_{c}(\varepsilon_{0},\alpha,\epsilon_\sigma)}_0 \frac{d\lambda}{\sqrt{\tilde{R}(\xi)}}, \label{jtabsmono}  \\
    J_{t}^{\mathrm{(scat)}}(\xi)&=& -2 A m_{0}^{3} \xi \sum_{\epsilon_{\sigma}=\pm1} \int^{\infty}_{\varepsilon_{\mathrm{min}}(\xi,\alpha,\epsilon_{\sigma})} d\varepsilon \delta(\varepsilon-\varepsilon_{0})\varepsilon \int_{\lambda_{c}(\varepsilon,\alpha,\epsilon_\sigma)}^{\lambda_{\mathrm{max}}(\xi,\varepsilon,\alpha,\epsilon_{\sigma})} \frac{d\lambda}{\sqrt{\tilde{R}(\xi)}} \nonumber \\
    &=& -2 A  m^3_{0} \xi \varepsilon_{0} \sum_{\epsilon_{\sigma}=\pm1} \Theta \left(\varepsilon_{0}-\varepsilon_{\mathrm{min}}(\xi,\alpha,\epsilon_{\sigma})\right) \int_{\lambda_{c}(\varepsilon_{0},\alpha,\epsilon_\sigma)}^{\lambda_{\mathrm{max}}(\xi,\varepsilon_{0},\alpha,\epsilon_{\sigma})} \frac{d\lambda}{\sqrt{\tilde{R}(\xi)}}, \label{jtscatmono} \\
    J_{\varphi}^{\mathrm{(abs)}}(\xi)&=&  A M m_{0}^{3} \xi \sum_{\epsilon_{\sigma}=\pm1} \int^{\infty}_{1} d\varepsilon \delta(\varepsilon-\varepsilon_{0}) \int^{\lambda_{c}(\varepsilon,\alpha,\epsilon_\sigma)}_0 d\lambda \frac{(\epsilon_{\sigma}\lambda+\alpha \varepsilon)}{\sqrt{\tilde{R}(\xi)}} \nonumber \\
    & = & A M m^3_{0} \xi \Theta(\varepsilon_{0}-1) \sum_{\epsilon_{\sigma}=\pm1}  \int^{\lambda_{c}(\varepsilon_{0},\alpha,\epsilon_\sigma)}_0 d\lambda \frac{(\epsilon_{\sigma}\lambda+\alpha \varepsilon_{0})}{\sqrt{\tilde{R}(\xi)}}, \label{jphiabsmono} \\
    J_{\varphi}^{\mathrm{(scat)}}(\xi)&=& 2 A M m_{0}^{3} \xi \sum_{\epsilon_{\sigma}=\pm1} \int^{\infty}_{\varepsilon_{\mathrm{min}}(\xi,\alpha,\epsilon_{\sigma})} d\varepsilon \delta(\varepsilon-\varepsilon_{0}) \int_{\lambda_{c}(\varepsilon,\alpha,\epsilon_\sigma)}^{\lambda_{\mathrm{max}}(\xi,\varepsilon,\alpha,\epsilon_{\sigma})} d \lambda \frac{(\epsilon_{\sigma}\lambda+\alpha \varepsilon_{0})}{\sqrt{\tilde{R}(\xi)}} \nonumber \\
    &=& 2 A M m_{0}^{3} \xi \sum_{\epsilon_{\sigma}=\pm1} \Theta \left (\varepsilon_{0}-\varepsilon_{\mathrm{min}}(\xi,\alpha,\epsilon_{\sigma})\right ) \int_{\lambda_{c}(\varepsilon_{0},\alpha,\epsilon_\sigma)}^{\lambda_{\mathrm{max}}(\xi,\varepsilon_{0},\alpha,\epsilon_{\sigma})} d\lambda \frac{(\epsilon_{\sigma}\lambda+\alpha \varepsilon_{0})}{\sqrt{\tilde{R}(\xi)}}, \label{jphiscatmono}  \\ 
    J^{r}_{\mathrm{(abs)}}(\xi)&=& -\frac{ A m^3_0}{\xi} \sum_{\epsilon_{\sigma}=\pm1} \int^\infty_1 d \varepsilon  \delta(\varepsilon-\varepsilon_{0}) \int^{\lambda_{c}(\varepsilon,\alpha,\epsilon_{\sigma})}_0 d\lambda = -\frac{ A m^3_0 }{\xi} \Theta \left (\varepsilon_{0}-1\right) \sum_{\epsilon_{\sigma}=\pm1} \lambda_{c}(\varepsilon_0,\alpha,\epsilon_{\sigma}), \label{jrabsmono} \\ 
    J^{r}_{\mathrm{(scat)}}(\xi)&=&0,
\end{eqnarray}
\end{subequations}
where $\Theta$ is the Heaviside step function. The additional factor 2 appearing in Eqs.\ (\ref{jtscatmono}) and (\ref{jphiscatmono}) is due to the incoming and outgoing trajectories, contributing to the total current. It is worth mentioning that the components of the particle current surface density $J_\mu^{(\mathrm{abs})}$ associated with the absorbed particles are regular at the horizon $\xi_{+}=1+\sqrt{1-\alpha^2}$. The contributions due to scattered particles (with fixed $\epsilon_\sigma = \pm 1$) vanish at the radius $\bar \xi$ satisfying the relation $\varepsilon_\mathrm{min}(\bar \xi,\alpha,\epsilon_\sigma) = \varepsilon_0$. Equation (\ref{eminlimit}) implies that $\xi_\mathrm{ph} < \bar \xi < \xi_\mathrm{mb}$.

In many applications, the proportionality constant $A$ can be hard to control. In such cases, it is convenient to introduce the asymptotic rest-mass surface density $\rho_{s,\infty}$, defined by
\begin{equation}
    \rho_{s,\infty} = m_0 n_{s,\infty}, \qquad n_{s,\infty} = \lim_{\xi \to \infty} \sqrt{-\gamma_{\mu \nu} J^\mu J^\nu},
\end{equation}
where $n_{s,\infty}$ denotes the asymptotic value of the particle number surface density. By performing asymptotic expansions (for $\xi \to \infty$) in Eqs.\ (\ref{monoenergeticanalyticresult}), one obtains\footnote{It is convenient to substitute $\lambda = \xi \tilde \lambda$.}
\begin{equation}
    \rho_{s,\infty} = - m_0 \lim_{\xi \to \infty} J_t = 2 \pi A m_0^4 \varepsilon_0 \Theta(\varepsilon_0 - 1).
\end{equation}

\subsection{Maxwell-J\"{u}ttner Type Distribution}

For the next model, we assume the distribution function $f$ in the form
\begin{equation}
\label{maxwellJuttner}
    f(x,p) = A\delta(\sqrt{-p_\mu p^\mu} -m_0)\exp\left(\beta p_t/m_0\right) = A \delta(m-m_0)\exp (-\beta \varepsilon)
\end{equation}
in the phase-space region populated by absorbed and scattered particles. As before, we set $f \equiv 0$ in the complementary region. Asymptotically (for $\xi \to +\infty$), Eq.\ (\ref{maxwellJuttner}) corresponds to a planar Maxwell-J\"{u}ttner distribution. The coefficient $\beta$ is related to the asymptotic temperature $T$ by $\beta = (m_0/k_\mathrm{B} T)$, where $k_\mathrm{B}$ is the Boltzmann constant.

In this case, explicit expressions for the absorbed and scattered components of $J_t$, $J_{\varphi}$ and $J^{r}$ can be written as
\begin{subequations}\label{jutterparticlesdensity}
\begin{eqnarray}
      J_{t}^{\text{(abs)}}(\xi)&=& - A  m_{0}^{3} \xi \sum_{\epsilon_{\sigma}=\pm1} \int^{\infty}_{1} d\varepsilon \exp(-\beta \varepsilon)\varepsilon \int^{\lambda_{c}(\varepsilon,\alpha,\epsilon_\sigma)}_0 \frac{d\lambda}{\sqrt{\tilde{R}(\xi)}}, \label{jtabsjuttner} \\
      J_{t}^{\text{(scat)}}(\xi)&=& -2 A m_{0}^{3} \xi \sum_{\epsilon_{\sigma}=\pm1} \int^{\infty}_{\varepsilon_{\mathrm{min}}(\xi,\alpha,\epsilon_{\sigma})} d\varepsilon \exp(-\beta \varepsilon)\varepsilon \int_{\lambda_{c}(\varepsilon,\alpha,\epsilon_\sigma)}^{\lambda_{\mathrm{max}}(\xi,\varepsilon,\alpha,\epsilon_{\sigma})} \frac{d\lambda}{\sqrt{\tilde{R}(\xi)}}, \label{jtscatjuttner}  \\
      J_{\varphi}^{\text{(abs)}}(\xi)&=&  A m_{0}^{3} M \xi \sum_{\epsilon_{\sigma}=\pm1} \int^{\infty}_{1} d\varepsilon \exp(-\beta \varepsilon) \int^{\lambda_{c}(\varepsilon,\alpha,\epsilon_\sigma)}_0 d\lambda \frac{(\epsilon_{\sigma}\lambda+\alpha \varepsilon)}{\sqrt{\tilde{R}(\xi)}}, \label{jphiabsjuttner} \\
       J_{\varphi}^{\text{(scat)}}&=& 2 A m_{0}^{3} M \xi \sum_{\epsilon_{\sigma}=\pm1} \int^{\infty}_{\varepsilon_{\mathrm{min}}(\xi,\alpha,\epsilon_{\sigma})} d\varepsilon \exp{(-\beta \varepsilon)} \int_{\lambda_{c}(\varepsilon,\alpha,\epsilon_\sigma)}^{\lambda_{\mathrm{max}}(\xi,\varepsilon,\alpha,\epsilon_{\sigma})}d\lambda\frac{(\epsilon_{\sigma}\lambda+\alpha \varepsilon)}{\sqrt{\tilde{R}(\xi)}}, \label{jphiscatjuttner} \\
        J^{r}_{\text{(abs)}}(\xi)&=-& \frac{ A m^3_0}{\xi} \sum_{\epsilon_{\sigma}=\pm1} \int^\infty_1 d\varepsilon \exp(-\beta \varepsilon) \lambda_{c}(\varepsilon,\alpha,\epsilon_\sigma), \label{jrabsjuttner} \\
       J^{r}_{\text{(scat)}}(\xi)&=&0.
\end{eqnarray}
\end{subequations}
As in the monoenergetic case, the factor 2 in Eqs.~\eqref{jtscatjuttner} and \eqref{jphiscatjuttner} is due to the contribution of the ingoing and outgoing trajectories.

The asymptotic rest-mass surface density can be computed as \cite{CMO2022} 
\begin{eqnarray}
    \rho_{s,\infty}\equiv m_0 n_{s,\infty}=-m_0 \lim_{\xi \to \infty} J_t = 2 \pi Am_0^4 \frac{1+\beta}{\beta^2}\exp{(-\beta)}.
\end{eqnarray}
Consequently, the factor $A$ in the integral expressions~(\ref{jutterparticlesdensity}) can be replaced by
\begin{eqnarray}
    A= \frac{    \rho_{s,\infty}}{2 \pi m^4_0}\frac{\beta^2}{1+\beta}\exp{(\beta)}.
\end{eqnarray}

\subsection{Angular velocity}

Some properties of kinetic planar accretion models, including accretion rates of the particle number, the energy, and the angular momentum, were analyzed in \cite{CMO2022} (in the Maxwell-J\"{u}ttner case) and in \cite{Khan2025} (for finite accretion disks). In this section, we add another element intended to illustrate the properties of the flow and define the averaged angular velocity, in the spirit of \cite{mms2025b}.

For the motion confined to the equatorial plane, one can define the mean particle four-velocity by
\begin{equation}
    u^\mu (x) = \frac{J^\mu(x)}{n_s(x)},
\end{equation}
where the particle surface density is defined in Eq.\ (\ref{ns})\footnote{Note that
\[ n_s = \sqrt{-J_t J^t - J_r J^r - J_\varphi J^\varphi}. \]
The components $J^t$, $J^\varphi$, and $J_r$ can be computed from $J_t$, $J_\varphi$, and $J^r$ by
\begin{eqnarray*}
J^t & = &\frac{1}{r^2 \Delta}\Big[2Mr\left(r^2J^r-aJ_\varphi\right)-\left(\Delta\left(r^2+2Mr\right)+4M^2r^2\right)J_t   \Big],\\
J^\varphi & = & \frac{1}{r^2 \Delta}\Big[\left(r^2-2Mr\right)J_\varphi+ar^2J^r+a\left(r^2+2Mr\right)J_t\Big], \\
J_r&=&\frac{1}{\Delta}\Big[r^2J^r-2MrJ_t-aJ_\varphi\Big].
\end{eqnarray*} 
}. The four-velocity is normalized as $\gamma_{\mu \nu} u^\mu u^\nu = -1$, as in standard general-relativistic hydrodynamics.

Killing vectors $k$ and $\chi$ allow one to introduce the specific energy $e$ and the azimuthal angular momentum $\ell_z$, defined by
\begin{eqnarray}
    e \coloneqq - k^\mu u_\mu =-\frac{J_t}{n_s}, \qquad \ell_z \coloneqq \eta^\mu  u_\mu =\frac{J_\varphi}{n_s}.
\end{eqnarray}
Both $e$ and $\ell_z$ transform as scalar quantities.

In the region outside the horizon, $r>r_+$, the vector field $J$ can be decomposed into two components: a linear combination of the Killing vectors $k$ and $\eta$ and a component orthogonal to the plane spanned by $k$ and $\eta$:
\begin{eqnarray}\label{angular_momentum111}
    J=N(k+\Omega\chi)+J^{\bot}.
\end{eqnarray}
Here, $N>0$, denotes a normalization constant, $J^{\bot}$ is orthogonal to $k$ and $\chi$, and $\Omega$ represents the angular velocity. The latter can be expressed as \cite{mms2025b}
\begin{equation}
    \Omega = \frac{\left(\chi^\mu J_\mu \right)\left( k^\nu k_\nu \right)-\left (k^\mu  J_\mu \right)\left(\chi^\nu  k_\nu \right)}{\left(k^\rho  J_\rho\right)\left(\chi^\sigma \chi_\sigma\right)-\left(\chi^\rho  J_\rho \right)\left(k^\sigma \chi_\sigma \right)} = \frac{\left(2M-r\right)J_{\varphi}+2aMJ_{t}}{\left(r^3+ra^2+2Ma^2\right)J_{t}+2aMJ_{\varphi}}.
\end{equation}
For $J_\varphi = 0$, $\Omega$ reduces to the angular velocity of a zero angular momentum observer (ZAMO)
\begin{eqnarray}
    \Omega_\mathrm{ZAMO}=\frac{2aM}{r^3+a^2\left(r+2M\right)}.
\end{eqnarray}

\begin{figure}
\subfigure[\label{omega12}]{\includegraphics[width=0.48\textwidth]{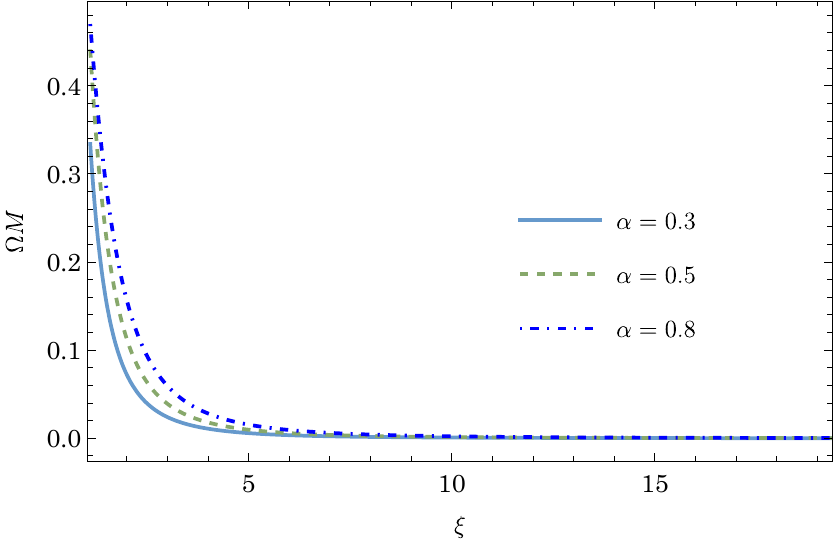}}
\subfigure[\label{angularenergy123}]{\includegraphics[width=0.49\textwidth]{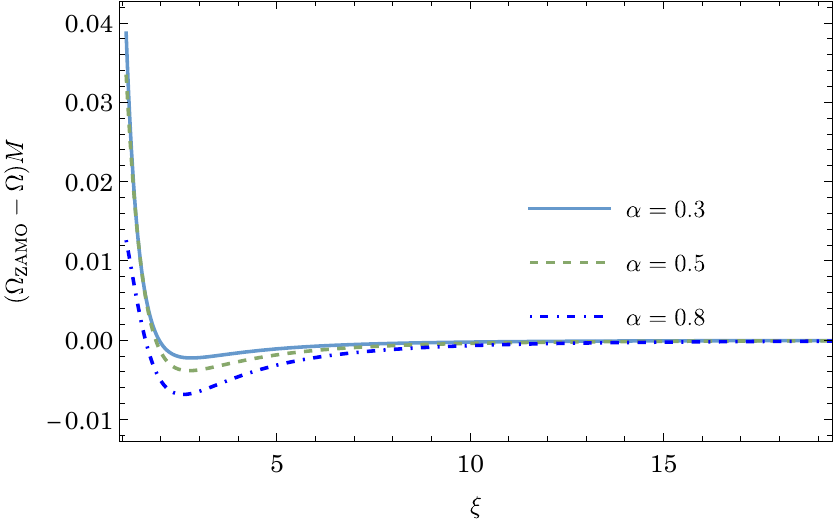}}
\subfigure[\label{angularmomentum12}]{\includegraphics[width=0.49\textwidth]{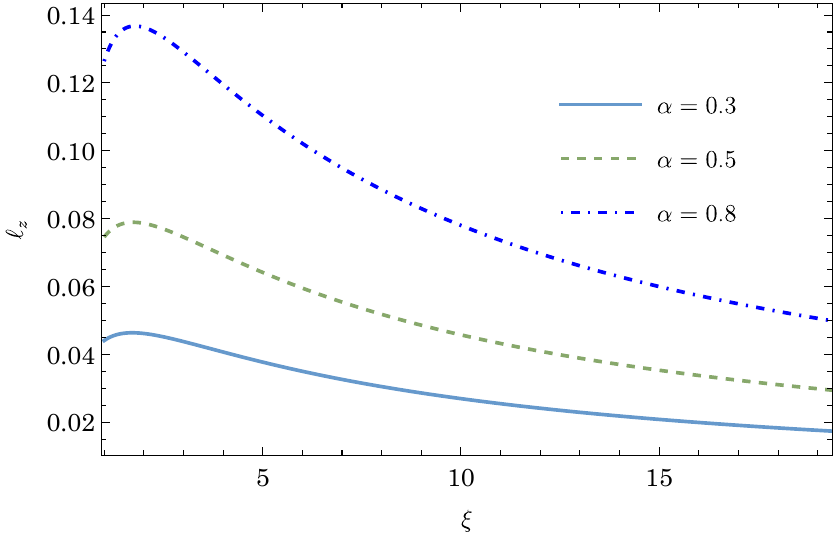}}
\subfigure[\label{angularenergy12}]{\includegraphics[width=0.49\textwidth]{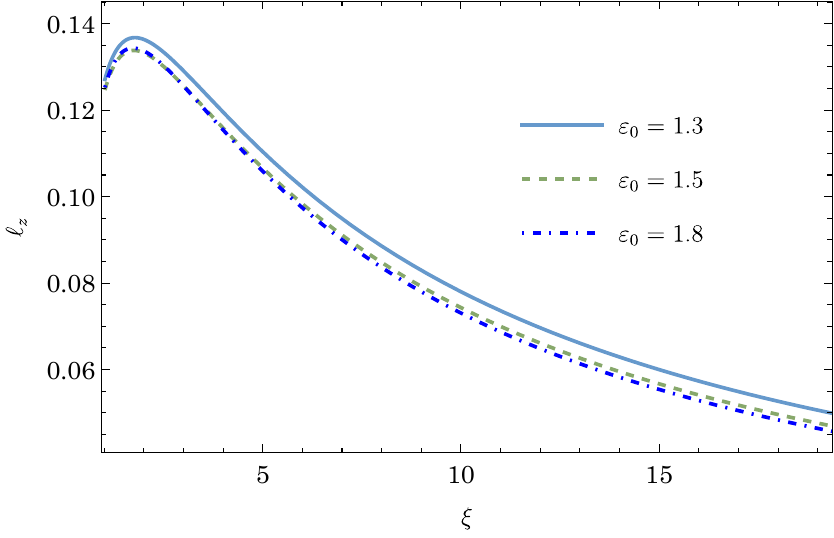}}
\caption{The angular velocity $\Omega$ and the azimuthal angular momentum $\ell_z$ for monoenergetic models. Subfigure \ref{omega12} illustrates the angular velocity $\Omega$, while Subfig.\ \ref{angularenergy123} shows the difference of $\Omega_{\mathrm{ZAMO}}-\Omega$. Subfig.\ \ref{angularmomentum12} shows the azimuthal angular momentum for three distinct values of the spin parameter $\alpha$. In both cases, the energy is fixed at $\varepsilon_0 = $1.3. In Subfigure \ref{angularenergy12} we plot the azimuthal angular momentum for three different values of the energy, with the black hole spin parameter set to $\alpha = 0.8$.}
\label{angularmomentum122}
\end{figure}

Sample plots of $\ell_z$ and $\Omega$ for monoenergetic models with $\alpha = 0.3$, 0.5, and 0.8 and $\varepsilon_0 = 1.3$, $1.5$, and $1.8$ are shown in Fig.\ \ref{angularmomentum122}. They show small but non-zero values of $\ell_z$ and angular velocities $\Omega$ deviating from $\Omega_\mathrm{ZAMO}$. This remains in agreement with Bondi-type solutions (not restricted to the equatorial plane) investigated in \cite{mms2025b}.

\section{Monte Carlo estimators}
\label{sec:montecarlo}

Our Monte Carlo approach can be characterized as an attempt to approximate a smooth distribution $\mathcal F$ function by a discrete one $\mathcal F^{(N)}$, supported at individual particle trajectories. The discrete distribution function $\mathcal F^{(N)}$ can be written as
\begin{eqnarray}\label{Montedist}
\mathcal F^{(N)}(x^\mu,p_\nu) = \sum_{i=1}^N \int \delta^{(4)}\left(x^{\mu}-x^{\mu}_{(i)}(\tau)\right)\delta^{(4)}\left(p_{\nu}-p_{\mu}^{(i)}(\tau)\right) d\tau.
\end{eqnarray}
Here, $N$ denotes the number of particles, each following a trajectory characterized by coordinates $x^\mu_{(i)}(\tau)$ and momenta $p_{\nu}^{(i)}(\tau)$. The affine parameter $\tau$ is defined as in Eqs.\ (\ref{eq:HamEqs}). Expression (\ref{Montedist}) can be regarded as a general-relativistic analog of a similar special-relativistic formula (\cite{Groot1980},~p. 14, Eq.~(A6)). The same expression appears in~\cite{Kampen1969}, differing by a factor $1/m$. The Monte Carlo scheme introduced in \cite{MCO2023, CMO2024} can be understood as a particular coarse-graining scheme, leading from a fine-grained distribution function $\mathcal F^{(N)}$ to a smooth one $\mathcal F$, adapted to the symmetries of the problem and to the numerical grid. Since, in practice,  one is only interested in observable quantities, such as the particle current density $\mathcal{J_\mu}$ or the energy-momentum tensor $\mathcal T_{\mu \nu}$, this procedure is only applied to observable quantities, in particular to $\mathcal{J_\mu}$. A general discussion of coarse-graining in the context of the special-relativistic kinetic theory is provided in~\cite{Kampen1969}. In view of Eq.~(\ref{Montedist}), the associated particle current density $\mathcal J_\mu^{(N)}(x)$ is given by
\begin{eqnarray}
\label{Jdiscrete}
    \mathcal J_\mu^{(N)}(x)&=&\int_{P_x^+} \mathcal F^{(N)}(x,p) p_\mu \sqrt{-\mathrm{det}g^{\alpha\beta}(x)} dp_0 dp_1 dp_2 dp_3 
    = \sum_{i=1}^N \int \delta^{(4)}\left(x^\alpha-x^\alpha_{(i)}(\tau)\right) p_\mu^{(i)}(\tau) \sqrt{-\mathrm{det}g^{\alpha\beta}(x)} d\tau,
\end{eqnarray}
where we have assumed that momenta associated with all trajectories are timelike and future-pointing. It can also be shown that the above expression agrees with Eq.~\eqref{particlecurrentdensity} (see \cite{MCO2023} for further details). 

Let $\Sigma \subset \mathcal M$ and $\sigma \subset \Sigma$ denote three-dimensional hypersurfaces embedded in $\mathcal M$. We think of $\sigma$ as a small numerical cell in $\Sigma$. The Monte Carlo estimator of $\mathcal J_\mu$ at a point $x \in \sigma$ is then defined as
\begin{equation}\label{montej}
    \langle \mathcal J_\mu(x) \rangle = \frac{\int_\sigma \mathcal J^{(N)}_\mu(x) \eta_\Sigma}{\int_\sigma  \eta_\Sigma},
\end{equation}
where $\eta_{\Sigma}$ denotes the volume element on $\Sigma$. For a stationary configuration, $\Sigma$ may be taken as a timelike hypersurface. In explicit examples, discussed in \cite{MCO2023}, computing the integrals appearing in Eq.\ (\ref{montej}) amounts to counting the intersections of particle trajectories with the hypersurfaces $\sigma$, with suitable weights.

In the planar case, relevant for this work, Monte Carlo estimators are defined for the particle current surface density $J_\mu$. We define
\begin{equation}
    \langle J_\mu(x) \rangle = \frac{\int_{\tilde \sigma} J^{(N)}_\mu(x) \eta_{\tilde \Sigma}}{\int_{\tilde \sigma}  \eta_{\tilde \Sigma}},
\end{equation}
where hypersurfaces $\tilde \Sigma$ and cells $\tilde \sigma$ are two-dimensional.

A specific example, chosen in this paper and adapted to the planar motion in the Kerr spacetime, is similar to the one of \cite{CMO2024}. It can be summarized as follows. We consider a surface of constant radius $r=r_0$ and constant $\vartheta = \pi/2$, defined by 
\begin{equation}
   \tilde \Sigma = \{(t,r,\vartheta,\varphi) \colon t \in \mathbb R, r=r_{0},\vartheta=\pi/2,  \varphi \in[0,2\pi)   \}.
\end{equation}
As a numerical cell $\tilde{\sigma}$ we take a subset of $\tilde \Sigma$, enclosed between two hypersurfaces of constant time $t_1$ and $t_2$:
\begin{eqnarray}
\label{setsigma}
    \tilde \sigma = \{(t,r,\vartheta,\varphi) \colon t_1\le t \le t_2, r=r_{0},\vartheta=\pi/2,  \varphi_{1}\le\varphi\le \varphi_{2}   \}.
\end{eqnarray}
This choice, adapted to the used coordinate system, reflects the stationarity of the spacetime. Note that $\tilde \sigma$ can be written as the image $\tilde \sigma = \Phi_{[t_1,t_2]}(S)$, where $\Phi_\tau(x_0^i)$ denotes the orbit of the Killing vector field $k = \partial_t$ passing through $x_0^i$ at $\tau = 0$. Here the set $S$ is defined as
\begin{eqnarray}
\label{setS}
S=\{(r,\vartheta,\varphi):r=r_{0},\vartheta=\pi/2,  \varphi_{1}\le\varphi\le \varphi_{2} \}.
\end{eqnarray}

The metric induced on $\tilde{\Sigma}$, can be written as
\begin{eqnarray}
     \tilde{g}&=&-\left(1-\frac{2M}{r_0}\right)dt^2-\frac{4 a M}{r_0} dt d\varphi + \left(r^2+a^2+\frac{2 a^2 M}{r_0}\right) d\varphi^2.
\end{eqnarray}
Consequently, the volume element on $\tilde \Sigma$ reads $\eta_{\tilde \Sigma}=\sqrt{-\mathrm{det}\tilde{g}_{\mu \nu}}dtd\varphi = \sqrt{\Delta_0} dt d\varphi$, where $\Delta_0 = r_0^2 - 2 M r_0 + a^2$.

For a planar model with trajectories confined to the equatorial plane, Eq.\ (\ref{Jdiscrete}) yields
\begin{eqnarray}
\label{currentN}
    \mathcal{J}_{\mu}^{(N)}(t,r,\theta,\varphi)= \sum^{N}_{i=1} \int \delta \left(\vartheta-\frac{\pi}{2}\right) \delta^{(3)}\left( x^{\alpha}-x^{\alpha}_{(i)} (\tau)
 \right)  p_\mu^{(i)} (\tau)  \sqrt{-\text{det}g^{\alpha\beta}(x)} d\tau,
\end{eqnarray}
with
\begin{eqnarray}
    \delta^{(3)}\left( x^{\alpha}-x^{\alpha}_{(i)} (\tau) \right)=\delta \left(t-t_{(i)} (\tau)\right) \delta\left(r_0-r_{(i)} (\tau)\right) \delta\left(\varphi-\varphi_{(i)} (\tau)\right).
\end{eqnarray}
The particle current surface density $J^{(N)}_\mu$ is defined, in analogy to Eq.\ (\ref{surfacedens}), as $\mathcal{J}_{\mu}^{(N)}=\delta(z)J_{\mu}^{(N)}$. Since in the equatorial plane $\delta\left(\vartheta-\frac{\pi}{2}\right)=r\delta(z)$, one can also write
\begin{eqnarray}
    \mathcal{J}_\mu^{(N)}(t,r,\vartheta,\varphi)=\frac{1}{r} \delta \left(\vartheta-\frac{\pi}{2}\right) J_\mu^{(N)}(t,r,\varphi),
\end{eqnarray}
and, in light of Eq.\ (\ref{currentN}),
\begin{eqnarray}
    J^{(N)}_{\mu}(t,r,\varphi) = r \sum^{N}_{i=1}\int \delta^{(3)}\left( x^{\alpha}-x^{\alpha}_{(i)} (\tau) \right)  p_\mu^{(i)} (\tau)  \sqrt{-\mathrm{det}g^{\alpha\beta}(x)} d\tau.
\end{eqnarray}

Integrating the components $J_\mu^\mathrm{(N)}$ over $\tilde \sigma$, one obtains
\begin{eqnarray}
    \int_{\tilde \sigma} J_{\mu}^{(N)} \eta_{\tilde \Sigma} & = & \frac{\sqrt{\Delta_0}}{r_0} \int^{t_2}_{t_1} dt  \int^{\varphi_2}_{\varphi_1} d\varphi \sum_{i=1}^N  \int d\tau   \delta^{(3)}\left(x^{\alpha}-x^{\alpha}_{(i)}(\tau)\right) p_{\mu}^{(i)}(\tau)        \label{integral1} \\ \nonumber 
    &=&  \frac{\sqrt{\Delta_0}}{r_0} \sum_{i\in I(\tilde{\sigma})} \int d\tau \delta\left(r_{0}-r_{(i)}(\tau)\right) p_{\mu}^{(i)}(\tau),
\end{eqnarray}
where $I(\tilde{\sigma}) \subseteq \{1,..., N \}$ denotes the set of indices corresponding to all trajectories that intersect $\tilde{\sigma}$. Changing the variables in the term $\delta(r_0 - r_{(i)})$, we get
\begin{eqnarray}\label{deltasum}
    \delta\left(r_{0}-r_{(i)}(\tau)\right)= \sum_{k} \frac{\delta(\tau-\tau_{k})}{\left|\frac{d r_{(i)}}{d\tau}\Big|_{\tau=\tau_{k}}\right|},
\end{eqnarray}
where the sum in Eq.\ (\ref{deltasum}) is taken over all intersections of the  $i$-th trajectory with $\tilde \sigma$ (note that a given trajectory can intersect $\tilde \sigma$ more than once). Note that  
\begin{eqnarray}
      \frac{d r_{(i)}}{d\tau}&=& p^{r}_{(i)} = \epsilon_{r,(i)}\frac{ \sqrt{R\left(E_{(i)},r_0,a,l_{(i)},\epsilon_{\sigma,(i)}\right)}}{r^2_{0}}.
\end{eqnarray}
Finally, Eq.~(\ref{integral1}) takes the form 
\begin{eqnarray}
\int_{\tilde \sigma} J_{\mu}^{(N)} \eta_{\tilde \Sigma} & = & \sqrt{\Delta_0} \sum_{i\in I(\tilde{\sigma})} \sum_k \int d \tau  \frac{\delta(\tau-\tau_{k})}{\left|\frac{d r_{(i)}}{d\tau}\Big|_{\tau=\tau_{k}}\right|} p_{\mu}^{(i)}(\tau) 
     = r_0 \sqrt{\Delta_0}  \sum_{i=1}^{N_\mathrm{int}}  \frac{p_{\mu}^{(i)}}{ \sqrt{R\left(E_{(i)},r_0,a,l_{(i)},\epsilon_{\sigma,(i)}\right)}},
\end{eqnarray}
where in the last expression the index $i$ numbers all $N_\mathrm{int}$ intersections of trajectories with $\tilde \sigma$.
The area of $\tilde \sigma$ reads
\begin{eqnarray}
    \int_{\tilde \sigma} \eta_{\tilde \Sigma}&=& \sqrt{\Delta_0}\int^{t_2}_{t_1} dt \int^{\varphi_{2}}_{\varphi_{1}} d\varphi = \sqrt{\Delta_0}(t_{2}-t_{1})(\varphi_{2}-\varphi_{1}).
\end{eqnarray}
Thus, the general expression for the Monte Carlo estimator of the particle current surface density in the equatorial plane can be written as
\begin{eqnarray}\label{montecarloestimator}
    \langle J_{\mu} \rangle & = &\frac{\int_{\tilde{\sigma}} J_{\mu} \eta_{\tilde{\Sigma}}}{\int_{\tilde{\sigma}} \eta_{\tilde{\Sigma}}}
    = \frac{1}{Mm(t_2-t_1)(\varphi_2-\varphi_1)} \sum_{i=1}^{N_\mathrm{int}}  \frac{p_{\mu}^{(i)} \xi_{0}}{\sqrt{\tilde{R}\left(\varepsilon_{(i)},\xi_0,\alpha,\lambda_{(i)},\epsilon_{\sigma,(i)}\right)}} .
\end{eqnarray}

A stationary distribution function $\mathcal F$ satisfies
\begin{equation}
    k^\mu \frac{\partial \mathcal F}{\partial x^\mu} - p_\alpha \frac{\partial k^\alpha}{\partial x^\mu} \frac{\partial \mathcal F}{\partial p_\mu} = 0,
\end{equation}
where $k$ denotes the timelike Killing vector \cite{CMO2024}. For $k = \partial_t$, this condition reduces to $\partial_t \mathcal F = 0$, which means that the distribution $\mathcal F$ remains invariant with respect to time translations. This allows one to replace a calculation of full (spacetime) particle trajectories by a computation of their projections on hypersurfaces of constant time. For planar models discussed in this paper, it is sufficient to compute the components $r_{(i)}(\tau)$,  $\varphi_{(i)}(\tau)$, $i = 1, \dots, N$, only. Similarly, instead of considering intersections of particle trajectories with $\tilde \sigma$, one can consider intersections with $S$, defined in Eq.\ (\ref{setS}). In this case, the normalization factor $t_2 - t_1$ appearing in Eq.\ (\ref{montecarloestimator}) has to be replaced by a term proportional to the total number of trajectories.

\section{Selection of geodesic parameters} \label{geodesic_parameters}

The selection of geodesic parameters is a crucial aspect of the Monte Carlo framework used in this paper. It should be emphasized that regardless of the sample of geodesic trajectories taken into account, the resulting Monte Carlo estimators computed according to Eq.\ (\ref{montej}) will always correspond to a valid solution of the Vlasov equation. A Monte Carlo approximation of an analytic solution can be obtained by selecting an appropriate set of particle trajectories. In the examples presented in this paper, we do this by selecting parameters of particle trajectories from the assumed asymptotic distribution.

Geodesic trajectories relevant for our model can be characterized by parameters: $\left\{ \xi_\mathrm{out}, \varphi_{0,(i)}, \varepsilon_{(i)}, \lambda_{(i)}, \epsilon_{\sigma,(i)} \right\}$, $i = 1, \dots, N$. As before, $\left\{ \varepsilon_{(i)}, \lambda_{(i)}, \epsilon_{\sigma,(i)} \right\}$ denote the constants of motion $\{ \varepsilon, \lambda, \epsilon_\sigma \}$ of the $i$-th trajectory. We assume that the $i$-th trajectory passes through the point with coordinates $(\xi,\varphi) = \left( \xi_\mathrm{out},\varphi_{0,(i)} \right)$, where it is characterized by $\epsilon_r = -1$. One could paraphrase this parametrization by saying that all particles are being ``injected'' from a circle of radius $\xi_\mathrm{out}$ towards the central black hole.

The Parameters $\left\{ \xi_\mathrm{out}, \varphi_{0,(i)}, \varepsilon_{(i)}, \lambda_{(i)}, \epsilon_{\sigma,(i)} \right\}$ should be selected in a way that reflects the desired distribution function; hence the procedure is different for the monoenergetic and the Maxwell-J\"{u}ttner models. For the monoenergetic model, we set $\varepsilon_{(i)} = \varepsilon_0$. Two samples, with $\epsilon_{\sigma,(i)} = +1$ and $\epsilon_{\sigma,(i)} = -1$ can be created separately. For each of these two samples, we select the values of $\lambda_{(i)}$ from a uniform distribution in the range $0 \le \lambda_{(i)} < \lambda_\mathrm{max}(\xi_\mathrm{out},\varepsilon_0,\alpha,\epsilon_{\sigma,(i)})$. All selected parameters can be subsequently divided into two classes, corresponding to absorbed $\left(0 \le \lambda_{(i)} < \lambda_c(\varepsilon_0,\alpha,\epsilon_{\sigma,(i)}) \right)$ and scattered $\left( \lambda_c(\varepsilon_0,\alpha,\epsilon_{\sigma,(i)})< \lambda_{(i)} \le \lambda_{\mathrm{max}}\left(\xi_\mathrm{out},\varepsilon_{0},\alpha,\epsilon_{\sigma,(i)}\right) \right)$ orbits. In principle, the values of the azimuthal angle $\varphi_{0,(i)} \in [0,2\pi)$ can be selected from a uniform distribution, together with the values of $\lambda_{(i)}$. On the other hand, since our models are axially symmetric, one can also choose $\varphi_{0,(i)} = 0$ for all $i = 1, \dots, N$ and set $\varphi_1 = 0$ and $\varphi_2 = 2\pi$, at the same time. 

For each of the two classes (with $\epsilon_{\sigma,(i)} = \pm 1$) the Monte Carlo estimators of $J_\mu$ can now be written as 
\begin{subequations}\label{Monoenergticestimator}
\begin{eqnarray}
\langle J_{t}^{(\mathrm{abs})} \rangle & = & -\frac{2 \pi  Am_{0}^{3} \lambda_{c}(\varepsilon_{0},\alpha,\epsilon_{\sigma})}{N_{\mathrm{abs}}(\varphi_2-\varphi_1)}  \sum_{i\in I_{\mathrm{abs}}(\varphi_{1},\varphi_{2})}  \frac{\varepsilon_{0} \xi_{0}}{ \sqrt{\tilde{R}\left(\xi_0,\varepsilon_{0},\alpha,\lambda_{(i)},\epsilon_{\sigma,(i)}\right)}}, \label{jtmontecarlomonoabs} \\
\langle J_{t}^{(\mathrm{scat})} \rangle &=&-\frac{4 \pi Am_{0}^{3} \left[ \lambda_\mathrm{max}(\xi_0,\varepsilon_0,\alpha,\epsilon_{\sigma})-\lambda_{c}(\varepsilon_{0},\alpha,\epsilon_{\sigma})\right]}{N_{\mathrm{scat}}(\varphi_2-\varphi_1)} \sum_{i\in I_{\mathrm{scat}}(\varphi_{1},\varphi_{2})}  \frac{\varepsilon_{0} \xi_{0}}{ \sqrt{\tilde{R}\left(\xi_0,\varepsilon_{0},\alpha,\lambda_{(i)},\epsilon_{\sigma,(i)}\right)}}, \label{jtmontecarlomonoscat} \\
\langle J_{\varphi}^{(\mathrm{abs})} \rangle & = & \frac{  2 \pi A m_{0}^{3} M \lambda_{c}(\varepsilon_{0},\alpha,\epsilon_{\sigma})}{ N_{\mathrm{abs}}(\varphi_2-\varphi_1)}  \sum_{i\in I_{\mathrm{abs}}(\varphi_{1},\varphi_{2})}  \frac{(\alpha \varepsilon_{0}+ \epsilon_{\sigma,(i)} \lambda_{(i)}) \xi_{0}}{ \sqrt{\tilde{R}\left(\xi_0,\varepsilon_{0},\alpha,\lambda_{(i)},\epsilon_{\sigma,(i)}\right)}}, \label{jphimontecarlomonoabs} \\
\langle J_{\varphi}^{(\mathrm{scat})} \rangle & = & \frac{4 \pi A m_{0}^{3} M \left[ \lambda_\mathrm{max}(\xi_0,\varepsilon_0,\alpha,\epsilon_{\sigma})- \lambda_{c}(\varepsilon_{0},\alpha,\epsilon_{\sigma})\right] }{ N_{\mathrm{scat}}(\varphi_2-\varphi_1)}  \sum_{i\in I_{\mathrm{scat}}(\varphi_{1},\varphi_{2})}  \frac{(\alpha \varepsilon_{0}+ \epsilon_{\sigma,(i)}  \lambda_{(i)}) \xi_{0}}{ \sqrt{\tilde{R}\left(\xi_0,\varepsilon_{0},\alpha,\lambda_{(i)},\epsilon_{\sigma,(i)}\right)}},\label{jphimontecarlomonoscat} \\
\langle J^{r}_{(\mathrm{abs})} \rangle & = & - \frac{  2 \pi A m_{0}^{3} \lambda_{c}(\varepsilon_{0},\alpha,\epsilon_{\sigma})}{N_{\mathrm{abs}}(\varphi_2-\varphi_1)} \sum_{i\in I_{\mathrm{abs}}(\varphi_{1},\varphi_{2})}  \frac{1}{\xi_{0}} = -  \frac{2 \pi  Am_{0}^{3} \lambda_{c}(\varepsilon_{0},\alpha,\epsilon_{\sigma})\# I_\mathrm{(abs)}(\varphi_1,\varphi_2)}{N_{\mathrm{abs}}(\varphi_2-\varphi_1) \xi_0}, \label{jrmontecarlomonoabs}
\end{eqnarray}
\end{subequations}
where $N_{\mathrm{abs}}$ and $N_{\mathrm{scat}}$ denote the number of absorbed and scattered orbits, respectively. The sets of indices $I_\mathrm{abs}(\varphi_1,\varphi_2)$ and $I_\mathrm{scat}(\varphi_1,\varphi_2)$ correspond, respectively, to the intersections of the absorbed and scattered trajectories with the segments $S$ (with constant dimensionless radius $\xi_0$), defined in Eq.\ (\ref{setS}). The number of absorbed trajectories intersecting the segment $S$ is denoted as $\# I_\mathrm{(abs)}(\varphi_1,\varphi_2)$. Note that setting $\varphi_1 = 0$ and $\varphi_2 = 2 \pi$, we obtain
\begin{equation}
    \langle J^{r}_{(\mathrm{abs})} \rangle = \frac{  Am_{0}^{3} \lambda_{c}(\varepsilon_{0},\alpha,\epsilon_{\sigma})}{\xi_0},
\end{equation}
which coincides with the corresponding expression in Eq.\ (\ref{jrabsmono}). Equations (\ref{Monoenergticestimator}) are obtained directly from Eq.\ (\ref{montecarloestimator}), by replacing the term $1/(M(t_2 - t_1))$ with $A m_0^3 V_\mathrm{abs}/N_\mathrm{abs}$ and $A m_0^3 V_\mathrm{scat}/N_\mathrm{scat}$, for absorbed and scattered trajectories, respectively. Here
\begin{equation}
    V_\mathrm{abs} = 2 \pi \lambda_c(\varepsilon_0,\alpha,\epsilon_\sigma), \qquad V_\mathrm{scat} = 2 \pi \left[ \lambda_\mathrm{max}(\xi_0,\varepsilon_0,\alpha,\epsilon_{\sigma})- \lambda_{c}(\varepsilon_{0},\alpha,\epsilon_{\sigma})\right]
\end{equation}
refer to ``volumes'' in the parameter space available for parameters $\lambda_{(i)}$ and $\varphi_{0,(i)}$ (hence the factor $2 \pi$ in expressions for $V_\mathrm{abs}$ and $V_\mathrm{scat}$). The factor $m_0^3$ can be recovered from the dimensional analysis---we recall that the constant $A$ has a dimension of $M^{-2}m_0^{-3}$.

The selection procedure of geodesic parameters for planar Maxwell-J\"{u}ttner models is slightly different. As before, we consider $\epsilon_{\sigma,(i)} = \pm 1$ separately. The energies $\varepsilon_{(i)}$ are sampled from the distribution (\ref{maxwellJuttner}), where we introduce a cutoff value $\varepsilon_\mathrm{cutoff}$. This can be done using von Neumann's rejection method, as described in \cite{MCO2023}. We select, assuming uniform distributions, the values of the energies $\varepsilon_{(i)} \in [1,\varepsilon_\mathrm{cutoff}]$, the angular momenta $\lambda_{(i)} \in \left[ 0, \lambda_\mathrm{max}\left(\xi_\mathrm{out},\varepsilon_\mathrm{cutoff},\alpha,\epsilon_{\sigma,(i)} \right) \right]$, and auxiliary parameters $y_{(i)} \in [0,1]$. These values $\left( \varepsilon_{(i)},\lambda_{(i)} \right)$ are included in the final set of parameters if they satisfy the conditions $y_{(i)} < \exp\left(-\beta\varepsilon_{(i)}\right)/\exp(-\beta)$,  $\lambda_{(i)}<\lambda_{\mathrm{max}}\left(\xi_\mathrm{out},\varepsilon_{(i)},\alpha,\epsilon_{\sigma,(i)}\right)$. Otherwise, they are rejected. As before, the values of $\varphi_{0,(i)}$ can be selected from the uniform distribution $\varphi_{0,(i)} \in [0,2\pi)$, or we can set $\varphi_{0,(i)} = 0$ and assume $\varphi_1 = 0$, $\varphi_2 = 2\pi$ in the definitions (\ref{setsigma}) and (\ref{setS}). Finally, parameters $\left\{ \xi_\mathrm{out}, \varphi_{0,(i)}, \varepsilon_{(i)}, \lambda_{(i)}, \epsilon_{\sigma,(i)} \right\}$ are divided into two groups, corresponding to absorbed and scattered orbits, depending on whether $\lambda_{(i)}$ is smaller or larger than $\lambda_c\left( \varepsilon_{(i)},\alpha,\epsilon_{\sigma,(i)} \right)$.

For each of the two classes (with $\epsilon_{\sigma,(i)} = \pm 1$) we write the Monte Carlo estimators of $J_\mu$ in the Maxwell-J\"{u}ttner model as
\begin{subequations}\label{jtmontejuttner}
\begin{eqnarray}
\langle J_{t}^{(\mathrm{abs})} \rangle &=&-\frac{ Am_{0}^{3} V_{\mathrm{abs}} }{N_{\mathrm{abs}}(\varphi_2-\varphi_1)}  \sum_{i\in I_{\mathrm{abs}}(\varphi_{1},\varphi_{2})}  \frac{\varepsilon_{(i)} \xi_{0}}{ \sqrt{\tilde{R}\left(\varepsilon_{(i)},\xi_0,\alpha,\lambda_{(i)},\epsilon_{\sigma,(i)}\right)}}, \label{jtmontecarlomonoabsju} \\
\langle J_{t}^{(\mathrm{scat})} \rangle & = & -\frac{ 2 Am_{0}^{3}V_{\mathrm{scat}}}{N_{\mathrm{scat}}(\varphi_2-\varphi_1)} \sum_{i\in I_{\mathrm{scat}}(\varphi_{1},\varphi_{2})}  \frac{\varepsilon_{(i)} \xi_{0}}{ \sqrt{\tilde{R}\left(\varepsilon_{(i)},\xi_0,\alpha,\lambda_{(i)},\epsilon_{\sigma,(i)}\right)}}, \label{jtmontecarlomonoscatju} \\
\langle J_{\varphi}^{(\mathrm{abs})} \rangle & = & \frac{ Am_{0}^{3} M V_{\mathrm{abs}}}{N_{\mathrm{abs}}(\varphi_2-\varphi_1)}  \sum_{i\in I_{\mathrm{abs}}(\varphi_{1},\varphi_{2})}  \frac{(\alpha \varepsilon_{(i)}+ \epsilon_{\sigma,(i)}  \lambda_{(i)}) \xi_{0}}{\sqrt{\tilde{R}\left(\varepsilon_{(i)},\xi_0,\alpha,\lambda_{(i)},\epsilon_{\sigma,(i)}\right)}}, \label{jphimontecarlomonosabsju} \\
\langle J_{\varphi}^{(\mathrm{scat})} \rangle & = & \frac{2 Am_{0}^{3} M V_{\mathrm{scat}}}{ N_{\mathrm{scat}}(\varphi_2-\varphi_1)}  \sum_{i\in I_{\mathrm{scat}}(\varphi_{1},\varphi_{2})}  \frac{(\alpha \varepsilon_{(i)}+\epsilon_{\sigma,(i)} \lambda_{(i)}) \xi_{0}}{ \sqrt{\tilde{R}\left(\varepsilon_{(i)},\xi_0,\alpha,\lambda_{(i)},\epsilon_{\sigma,(i)}\right)}},\label{jphimontecarlomonosscatju} \\
 \langle J^{r}_{(\mathrm{abs})} \rangle & = & - \frac{ Am_{0}^{3}V_{\mathrm{abs}}}{N_{\mathrm{abs}}(\varphi_2-\varphi_1)} \sum_{i\in I_{\mathrm{abs}}(\varphi_{1},\varphi_{2})}  \frac{1}{\xi_{0}} =-\frac{ Am_{0}^{3}V_{\mathrm{abs}}\# I_{\mathrm{abs}}(\varphi_{1},\varphi_{2})}{N_{\mathrm{abs}}(\varphi_2-\varphi_1)\xi_0}, \label{jrmontecarlomonosabsju}
\end{eqnarray}
\end{subequations}
where 
\begin{eqnarray}
V_{\mathrm{abs}}&=& 2 \pi \int^{\varepsilon_{\mathrm{cutoff}}}_1 d\varepsilon \exp{(-\beta \varepsilon)} \lambda_{c}(\varepsilon,\alpha,\epsilon_{\sigma}), \\
    V_{\mathrm{scat}}&=& 2 \pi\int^{\varepsilon_{\mathrm{cutoff}}}_1 d\varepsilon \exp{(-\beta \varepsilon)} \left[ \lambda_{\mathrm{max}}(\xi_\mathrm{out},\varepsilon,\alpha,\epsilon_{\sigma})-\lambda_{c}(\varepsilon,\alpha,\epsilon_{\sigma})\right].
\end{eqnarray}
The symbols $N_{\mathrm{abs}}$, $N_{\mathrm{scat}}$, $I_{\mathrm{abs}}(\varphi_1,\varphi_2)$, and $I_{\mathrm{scat}}(\varphi_1,\varphi_2)$ have the same meaning, as in Eqs.\ (\ref{Monoenergticestimator}). In analogy to the monoenergetic model, the expressions $V_\mathrm{abs}$ and $V_\mathrm{scat}$ refer to ``volumes'' in the parameter space of $\varepsilon_{(i)}$, $\lambda_{(i)}$, and $\varphi_{0,(i)}$.

\section{Numerical results}
\label{sec:numresults}

\begin{figure}[t]
\subfigure[\label{jtplus}]{\includegraphics[width=0.45\textwidth]{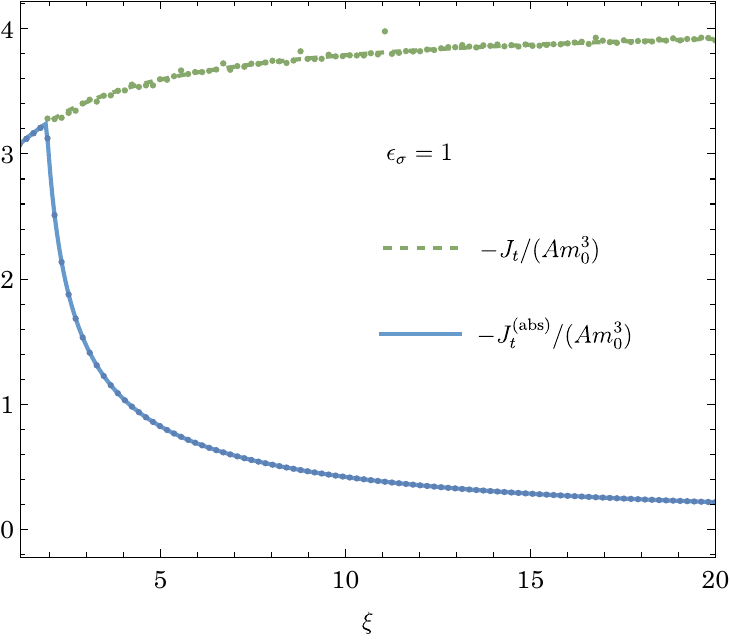}}
\subfigure[\label{jtminus}]{\includegraphics[width=0.45\textwidth]{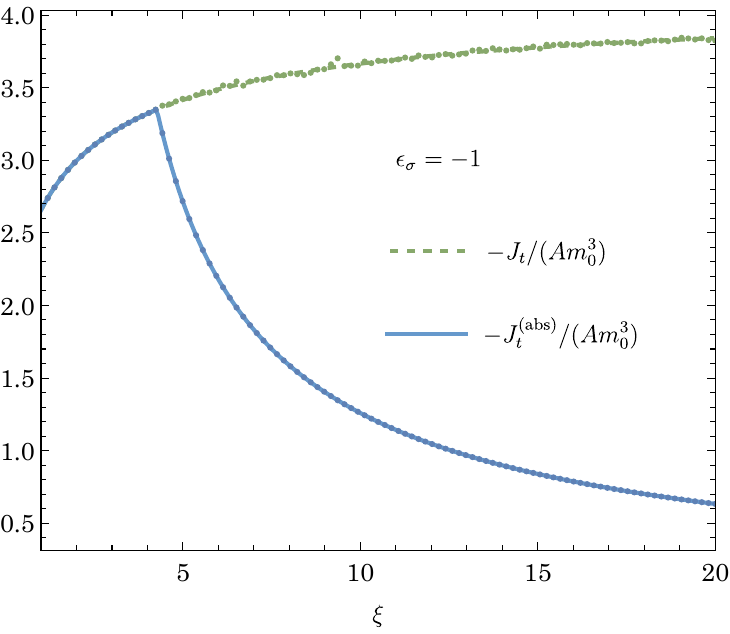}}
\caption{Time components of particle current surface density $J_{t}$ for the monoenergetic planar model with $\alpha = 0.8$, $\varepsilon_{0} = 1.3$, and $\xi_{\mathrm{out}} = 20$. Left and right panels show the results for $\epsilon_\sigma = +1$ and $\epsilon_\sigma = -1$, respectively. Exact solutions~\eqref{jtabsmono}--\eqref{jtscatmono} are plotted with solid and dashed lines. Dots (blue and green) represent Monte Carlo estimators \eqref{jtmontecarlomonoabs}--\eqref{jtmontecarlomonoscat}. A total of $10^{6}$ trajectories is simulated: in Subfig.\ \ref{jtplus}, $N_{\mathrm{abs}}$ = 166847, $N_{\mathrm{scat}}$ = 833153; in Subfig.\ \ref{jtminus}, $N_{\mathrm{abs}}$ = 424080, $N_{\mathrm{scat}}$ = 575920.}
\label{jtfigure}
\end{figure}

\begin{figure}[t]
\subfigure[\label{jphiplus}]{\includegraphics[width=0.45\textwidth]{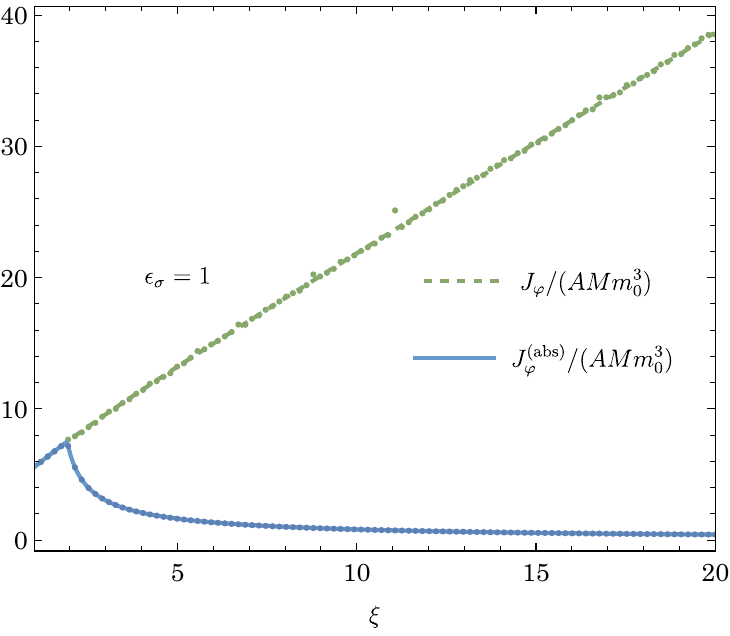}}
\subfigure[\label{jphiminus}]{\includegraphics[width=0.45\textwidth]{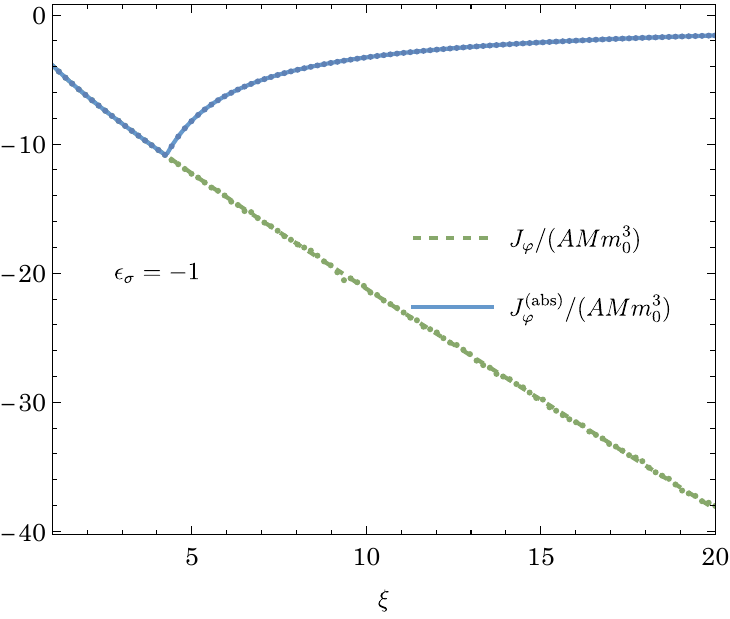}}
\caption{Angular components of particle current surface density $J_{\varphi}$ for the monoenergetic planar model with $\alpha = 0.8$, $\varepsilon_{0} = 1.3$, and $\xi_{\mathrm{out}} = 20$. Left and right panels show the results for $\epsilon_\sigma = +1$ and $\epsilon_\sigma = -1$, respectively. Exact solutions \eqref{jphiabsmono}--\eqref{jphiscatmono} are plotted with solid and dashed lines. Dots (blue and green) represent Monte Carlo estimators \eqref{jphimontecarlomonoabs}--\eqref{jphimontecarlomonoscat}. A total of $10^{6}$ trajectories is simulated: in Subfig.~\ref{jphiplus}, $N_{\text{abs}} $= 166847, $N_{\mathrm{scat}}$ = 833153; in Subfig.~\ref{jphiminus}, $N_{\mathrm{abs}}$ = 424080, $N_{\mathrm{scat}}$ = 575920.}
\label{jphifigure}
\end{figure}

\begin{figure}[t]
\subfigure{\includegraphics[width=0.45\textwidth]{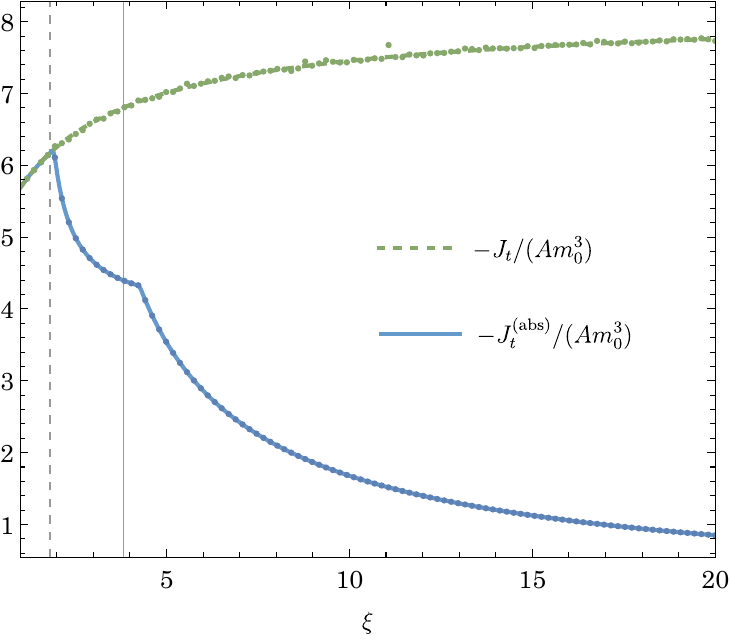}}
\subfigure{\includegraphics[width=0.45\textwidth]{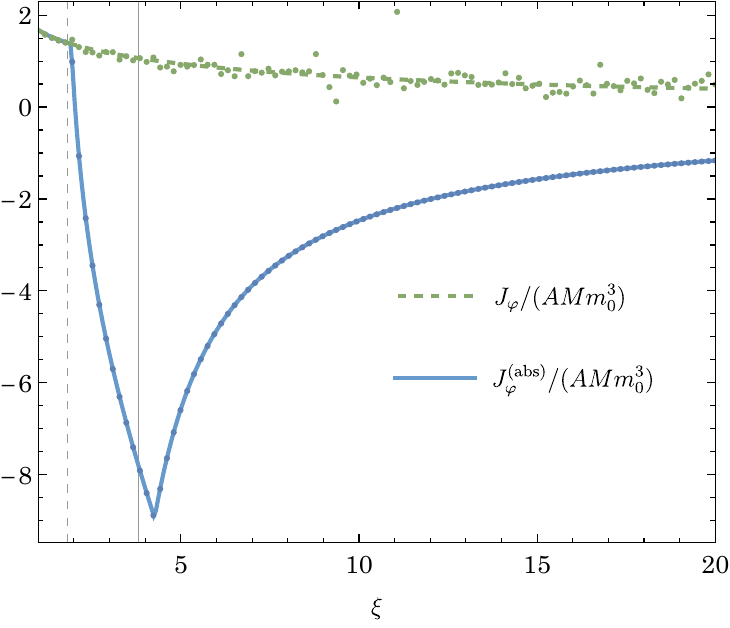}}
\caption{Total components (the sums of contributions with $\epsilon_\sigma = \pm1$) of the particle current surface density $J_t$ (left) and $J_\varphi$ (right). Both panels correspond to the planar monoenergetic model with $\alpha = 0.8$, $\varepsilon_{0} = 1.3$, and $\xi_{\mathrm{out}} = 20$. Vertical lines mark locations of circular photon orbits $\xi_{\mathrm{ph}}$ for $\epsilon_\sigma = \pm 1$.}
\label{j_t_current_density}
\end{figure}

\begin{figure}[t]
\subfigure[\label{jtjuttnerplus}]{\includegraphics[width=0.45\textwidth]{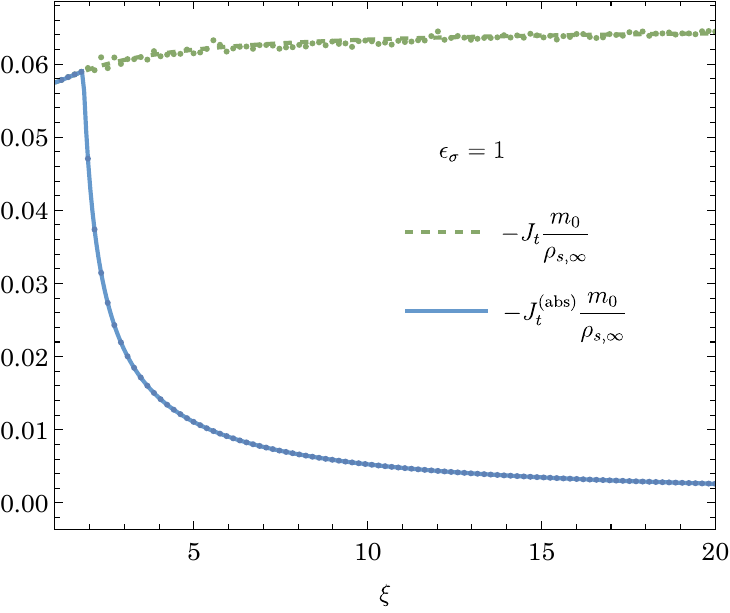}}
\subfigure[\label{jtjuttnerminus}]{\includegraphics[width=0.45\textwidth]{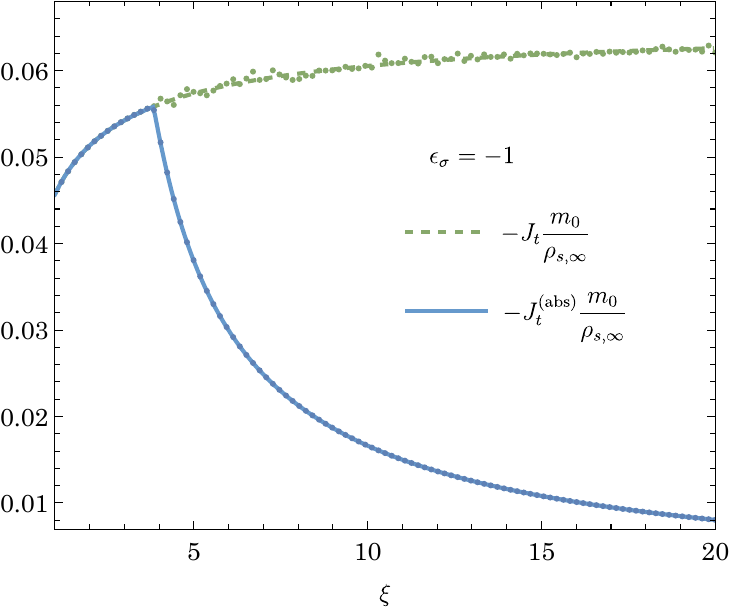}}
\caption{Time components of the particle current surface density $J_{t}$ for the Maxwell-J\"{u}ttner planar model with $\varepsilon_{\text{cutoff}}$ = 10, $\xi_{\mathrm{out}}$ = 20, $\beta=1/10$, and $\alpha$ = 0.8. Left and right panels show the results for $\epsilon_\sigma = +1$ and $\epsilon_\sigma = -1$, respectively. Exact solutions~\eqref{jtabsjuttner}--\eqref{jtscatjuttner} are plotted with solid and dashed lines. Dots (blue and green) represent Monte Carlo estimators \eqref{jtmontecarlomonoabsju}--\eqref{jtmontecarlomonoscatju}. In Subfigure \ref{jtjuttnerplus}, $N_{\mathrm{abs}}$ = 32298, $N_{\mathrm{scat}}$ = 272564; in Subfig.~\ref{jtjuttnerminus}, $N_{\mathrm{abs}}$ = 106663, $N_{\mathrm{scat}}$ = 203118.}
\label{jtfigurejuttner}
\end{figure}

\begin{figure}[t]
\subfigure[\label{jphijuttnerplus}]{\includegraphics[width=0.45\textwidth]{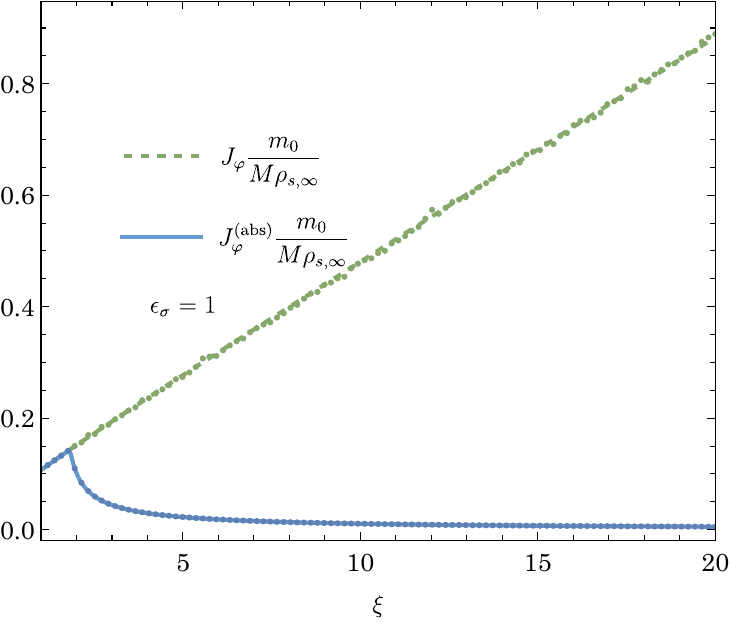}}
\subfigure[\label{jphijuttnerminus}]{\includegraphics[width=0.45\textwidth]{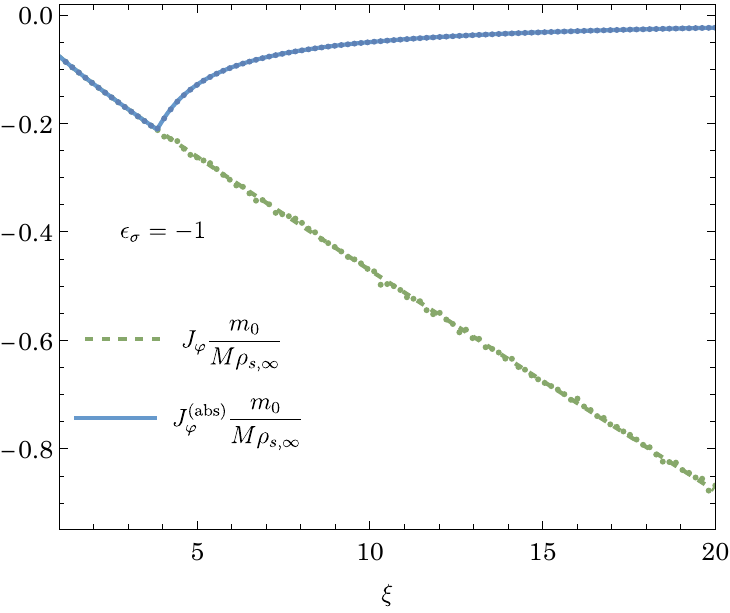}}
\caption{Angular components of particle current surface density $J_{\varphi}$ for the Maxwell-J\"{u}ttner planar model with $\varepsilon_{\text{cutoff}}$ = 10, $\xi_{\mathrm{out}}$ = 20, $\beta=1/10$, and $\alpha$ = 0.8. Left and right panels show the results for $\epsilon_\sigma = +1$ and $\epsilon_\sigma = -1$, respectively. Exact solutions \eqref{jphiabsjuttner}--\eqref{jphiscatjuttner} are plotted with solid and dashed lines. Dots (blue and green) represent Monte Carlo estimators \eqref{jphimontecarlomonosabsju}--\eqref{jphimontecarlomonosscatju}. In Subfigure \ref{jphijuttnerplus}, $N_{\text{abs}}$ = 32298, $N_{\mathrm{scat}}$ = 272564; in Subfig.~\ref{jphijuttnerminus}, $N_{\mathrm{abs}}$ = 106663, $N_{\mathrm{scat}}$ = 203118. }
\label{jphifigurejuttner}
\end{figure}

\begin{figure}[t]
\subfigure{\includegraphics[width=0.45\textwidth]{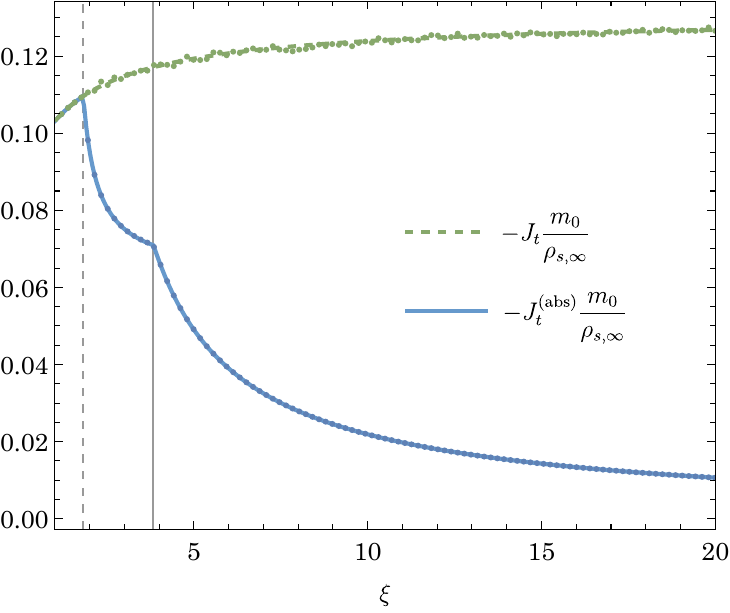}}
\subfigure{\includegraphics[width=0.45\textwidth]{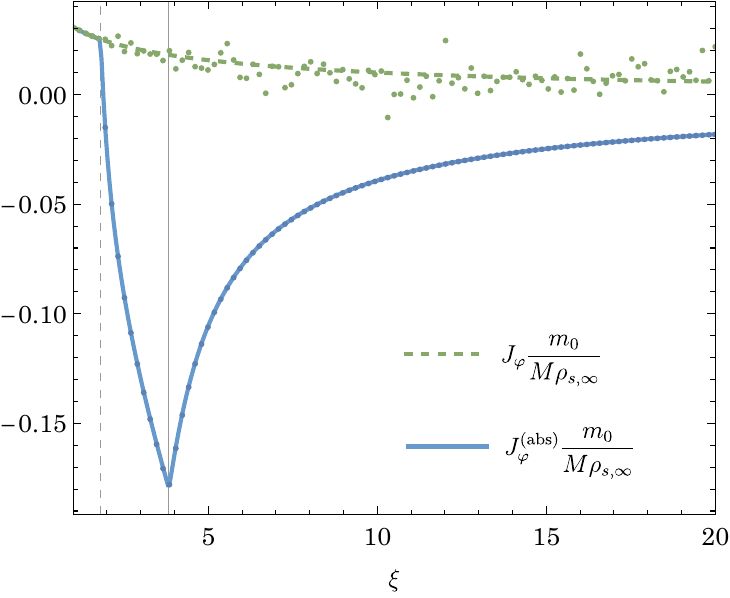}}
\caption{Total components (the sums of contributions with $\epsilon_\sigma = \pm1$) of the particle current surface density $J_t$ (left) and $J_\varphi$ (right). Both panels correspond to the planar Maxwell-J\"{u}ttner model with $\varepsilon_{\mathrm{cutoff}} = 10$, $\xi_{\mathrm{out}} = 20$, $\beta=1/10$, and $\alpha = 0.8$. Vertical lines mark locations of circular photon orbits $\xi_{\mathrm{ph}}$ for $\epsilon_\sigma = \pm 1$.}
\label{jtscatabscurrentdensity}
\end{figure}

In our numerical implementation, we compute the Monte Carlo estimators $\langle J_\mu \rangle$, derived in Sec.\ \ref{geodesic_parameters}. The employed algorithm is similar to the one used in \cite{MCO2023,CMO2024}. All numerical results shown in this work were obtained using the \textit{Wolfram Mathematica} \cite{WR2023}.

The numerical grid spans the radial range $1 < \xi < \xi_\mathrm{out}$, where $\xi_\mathrm{out}$ denotes the outer radius of the grid. Due to the assumed axial symmetry, we select the surfaces (\ref{setS}) with $\varphi_1 = 0$ and $\varphi_2 = 2 \pi$ (circles), assuming evenly distributed radii (denoted as $r_0 = M \xi_0$ in Eq.\ (\ref{setS})):
\begin{eqnarray}
    \xi_j= 1 + j\Delta\xi, \quad j=1, \dots, N_\xi,
\end{eqnarray}
where $\Delta\xi=(\xi_\mathrm{out} - 1)/N_\xi$. For the results presented in Figs.\ \ref{jtfigure}--\ref{jtscatabscurrentdensity}, we set $N_\xi$ = 100 and $\xi_\mathrm{out}$ = 20. Note that the choice of $\xi_\mathrm{out}$ does not influence the results of the numerical scheme (except affecting the radial resolution of the grid). The assumed distribution of geodesic parameters corresponds to a uniform gas at infinity, and only unbound orbits are taken into account. An analysis of kinetic accretion occurring from a finite distance in the equatorial plane of the Kerr spacetime was recently carried out in \cite{Khan2025}. It involves, among others, taking into account bound orbits that can reach the assumed outer edge of the disk.

As a consequence of our choice of sets $S$ as circles of constant $\xi = \xi_0$, every unbound absorbed trajectory intersects each circle $S$ only once. In contrast to that, every scattered trajectory intersects every circle $S$ twice, once (in the event that the radius of the circle coincides with the radial turning point), or not at all. If a scattered trajectory intersects the circle $S$ twice, one of the intersection points corresponds to $\epsilon_r = -1$ (ingoing particle) and the other to $\epsilon_r = +1$ (outgoing particle).

Since, due to axial symmetry, we set $\varphi_1 = 0$ and $\varphi_2 = 2 \pi$, we also choose the ``initial'' azimuthal angle $\varphi_{0,(i)} = 0$ for all trajectories. As a consequence, the exact expressions for $\xi(s)$ and $\varphi(s)$, and $s(\xi)$ (Eqs.\ (\ref{WeierstrassFormula}), (\ref{phiquadraturesc}), (\ref{S1}), and (\ref{S2})) are not actually used in the calculation. The expressions for $\varphi(s)$ and $s(\xi)$ (Eqs.\ (\ref{phiquadraturesc}), (\ref{S1}), and (\ref{S2})) would be necessary if we chose to discretize the azimuthal angle $\varphi$ as well. The latter option could be used in the analysis of solutions which break axial symmetry, similar to the case of a moving black hole, considered in \cite{CMO2024}. On the other hand, in the computation of components corresponding to scattered trajectories, we need to know the location of the pericenter of every trajectory in the sample, to establish the number of intersections with each of the circles $S$. The location of the pericenter can easily be computed as the largest root of $\tilde R(\xi)$.

\begin{figure}[t]
\subfigure[\label{orbits1}]{\includegraphics[width=0.45\textwidth]{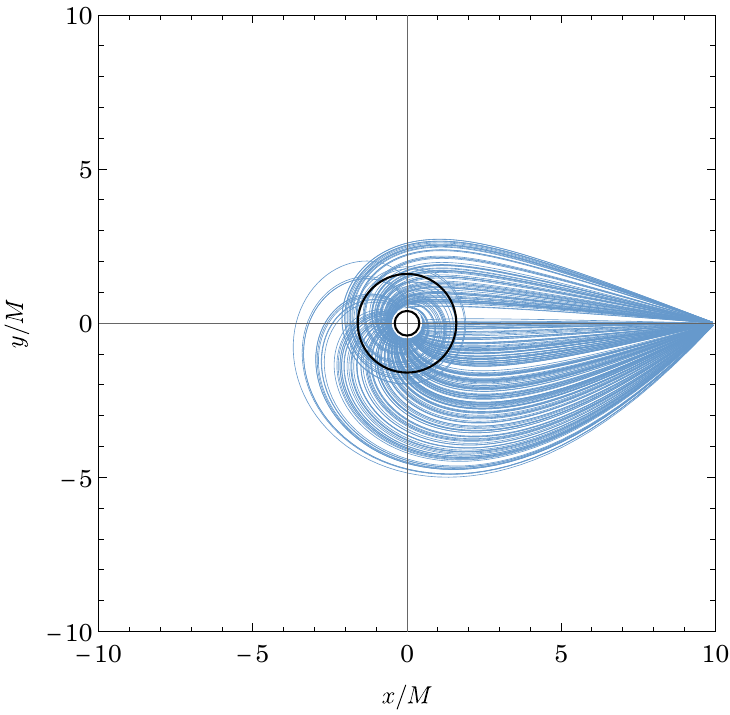}}
\subfigure[\label{orbits2}]{\includegraphics[width=0.45\textwidth]{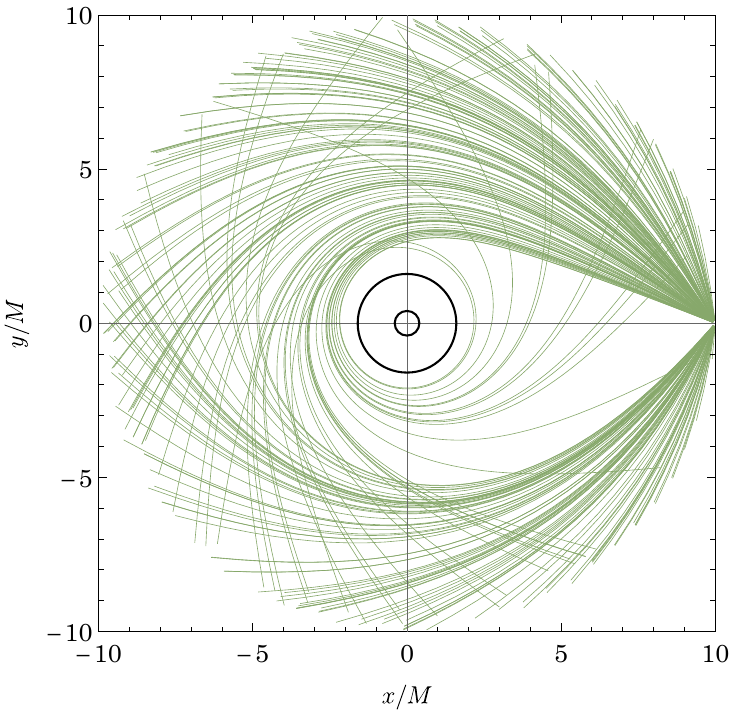}}
\subfigure[\label{orbits3}]{\includegraphics[width=0.45\textwidth]{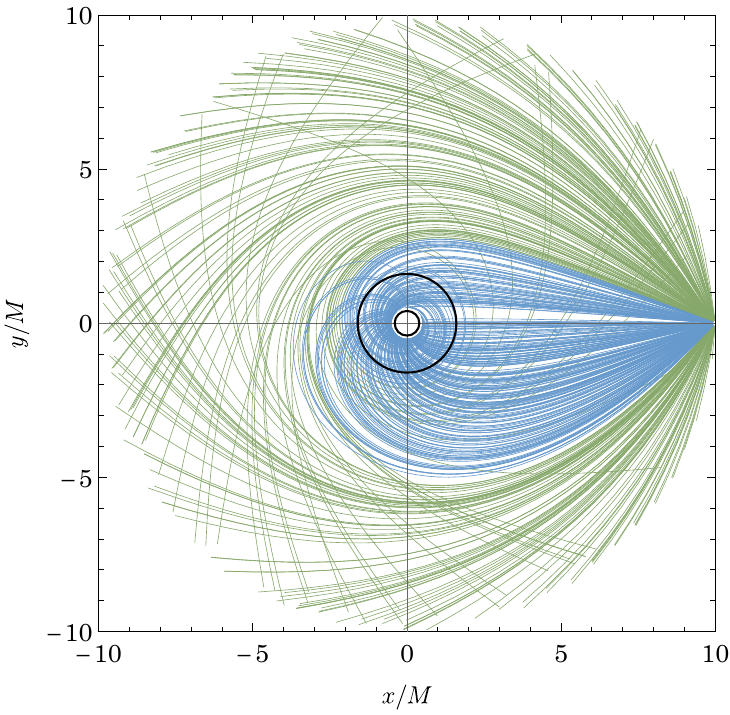}}
\subfigure[\label{orbits4}]{\includegraphics[width=0.45\textwidth]{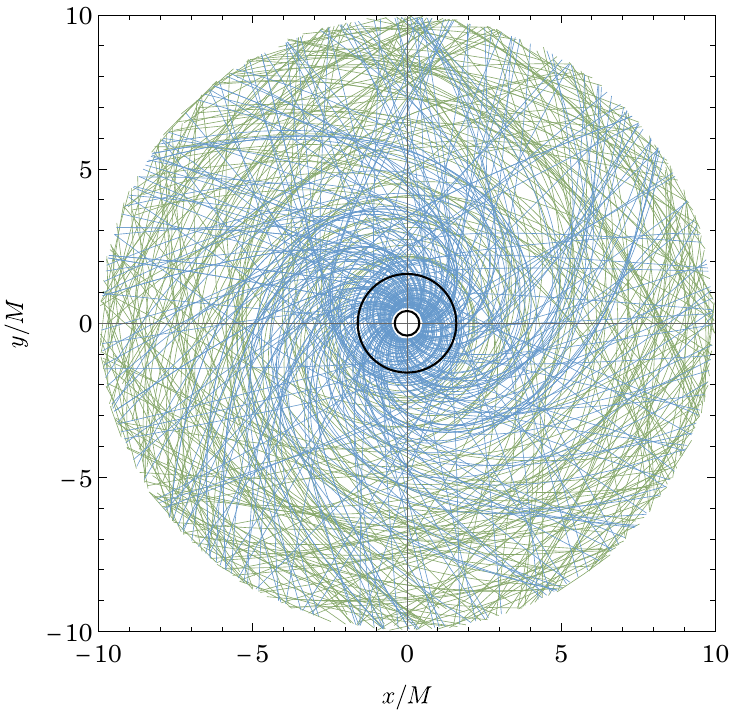}}
\caption{A monoenergetic random distribution of equatorial geodesic orbits in the Kerr spacetime for the parameters $\alpha = 0.8$, $\varepsilon_{0} = 1.3$, and $\xi_\mathrm{out}$ = 10. Absorbed and scattered geodesics are plotted with blue and green lines, respectively. Subfigures \ref{orbits1}--\ref{orbits3} illustrate the setup used in numerical calculations of Figs.\ \ref{jtfigure}--\ref{jtscatabscurrentdensity}, with $\varphi_{0,(i)} = 0$. In Subfigure \ref{orbits4}, the values of $\varphi_{0,(i)}$ are uniformly distributed in the range $0 \le \varphi_{0,(i)} < 2 \pi$. For clarity, a total of 500 trajectories is shown. Solid black circles mark the locations of the inner and outer Kerr horizons. \label{TrajectoryPlot}}
\end{figure}

Knowledge of all $\xi(s)$ and $\varphi(s)$, and $s(\xi)$ is required to plot the orbits of individual particle trajectories, shown in Fig.\ \ref{TrajectoryPlot}. In this case $s(\xi)$ is needed to control the range of the Mino time corresponding to the motion from $\xi_\mathrm{out}$ to the horizon, for absorbed trajectories, and from $\xi_\mathrm{out}$ to the pericenter and back to $\xi_\mathrm{out}$, for scattered orbits. 

Sample plots of the Monte Carlo estimators $\langle J_t \rangle$ and $\langle J_\varphi \rangle$ are shown in Figs.\ \ref{jtfigure}--\ref{jtscatabscurrentdensity}. Results for a monoenergetic model with $\varepsilon_0 = 1.3$ are shown in Figs.\ \ref{jtfigure}--\ref{j_t_current_density}. Figures \ref{jtfigurejuttner}--\ref{jtscatabscurrentdensity} illustrate a Maxwell-J\"{u}ttner model with $\beta = 1/10$. In all cases, Monte Carlo results are plotted using dots. Solid or dashed lines correspond to exact values, computed using expressions (\ref{monoenergeticanalyticresult}) and (\ref{jutterparticlesdensity}). To provide a fair comparison for the Maxwell-J\"{u}ttner model, when computing the exact values (\ref{jutterparticlesdensity}), we limit the energies $\varepsilon$ to the cutoff value ($\varepsilon_\mathrm{cutoff} = 10$) assumed in the selection of orbital parameters in the Monte Carlo scheme. We omit the plots of the radial components of the particle current surface density $J^r$, as they show a simple $1/\xi$ behavior. In all cases, we observe an excellent agreement between exact and Monte Carlo results.

In all Figures \ref{jtfigure}--\ref{jtscatabscurrentdensity}, blue and green colors are used for contributions associated with absorbed orbits and the sums of contributions corresponding to absorbed and scattered orbits, respectively.

Figures \ref{jtfigure}, \ref{jphifigure}, \ref{jtfigurejuttner}, and \ref{jphifigurejuttner} illustrate the components of the particle current surface density with fixed $\epsilon_\sigma = \pm 1$. A division of possible values of the azimuthal angular momentum $\lambda_z$ into $\epsilon_\sigma = \pm 1$ is, to some extent, arbitrary, and it is related to the convention assumed in Eq.\ (\ref{epssigmaconvention}). However, these figures allow us to show the locations in which the components of $J^\mathrm{(abs)}_t$, $J^\mathrm{(scat)}_t$,  $J^\mathrm{(abs)}_\varphi$,  and $J^\mathrm{(scat)}_\varphi$ lose smoothness (with respect to $\xi$). For components corresponding to scattered orbits, this behavior is related to the limit of $\varepsilon_\mathrm{min}$, given by Eq.\ (\ref{eminlimit}), which in turn depends on $\epsilon_\sigma$. In the monoenergetic model, the components of $J_\mu$ associated with scattered orbits vanish for $\xi$ smaller than the root of the equation $\varepsilon_\mathrm{min}(\xi,\alpha,\epsilon_\sigma) = \varepsilon_0$, which lies between $\xi_\mathrm{ph}$ and $\xi_\mathrm{mb}$. In the Maxwell-J\"{u}ttner model, the components $J_\mu^\mathrm{(scat)}$ vanish for $\xi < \xi_\mathrm{ph}$, which is easy to understand, as the solution of the equation $\varepsilon_\mathrm{min}(\xi,\alpha,\epsilon_\sigma) = \varepsilon_0$ tends to $\xi_\mathrm{ph}$ for $\varepsilon_0 \to \infty$.

It can be proved that the total currents---the sums of components associated with absorbed and scattered orbits---remain smooth functions of $\xi$ (see, e.g., Appendix D in \cite{mms2025b}). This behavior is also visible in all plots \ref{jtfigure}--\ref{jtscatabscurrentdensity}.

As a final example we plot, in Fig.\ \ref{TrajectoryPlot},  a monoenergetic selection of orbits in the equatorial plane $(x,y)$, where
\begin{eqnarray}
    x= r \cos{\varphi}, \qquad y = r \sin{\varphi}.
\end{eqnarray}
The absorbed and scattered trajectories are plotted with green and blue colors, respectively. All trajectories shown in Fig.\ \ref{TrajectoryPlot} originate at $\varphi_{0,(i)} = 0$, which corresponds to the choice used in our Monte Carlo scheme. For the purpose of this plot, the azimuthal angular momenta $\lambda_z$ are sampled uniformly in the range $\left(-\lambda_\mathrm{max}\left(\xi_\mathrm{out},\varepsilon_0,\epsilon_\sigma = -1\right) + \alpha \varepsilon_0, \lambda_\mathrm{max}\left( \xi_\mathrm{out},\varepsilon_0,\epsilon_\sigma = +1\right) + \alpha \varepsilon_0\right)$, as follows from Eqs.\ (\ref{epssigmaconvention}) and (\ref{lambdamax}).

\section{Discussion}
\label{sec:discussion}

We have extended a recent Monte Carlo scheme for computing stationary solutions of the general-relativistic Vlasov equation to the Kerr spacetime. In particular examples, we considered razor-thin accretion disks confined to the equatorial plane. This model, introduced in \cite{CMO2022}, was initially intended to approximate the Bondi-type accretion of a collisionless fluid in the Kerr spacetime, analyzed later in \cite{mms2025a,mms2025b}. The properties of the planar model \cite{CMO2022} and Bondi-type flows analyzed in \cite{mms2025a,mms2025b} differ, mostly due to the effective dimensionality of the asymptotic distribution of the gas. To some extent, this drawback of the infinite planar disk model of \cite{CMO2022} was remedied in \cite{Khan2025} by considering accretion occurring from a finite distance. This involves taking into account particles moving on bound orbits that can meet the outer edge of the disk.

Despite the above criticism, the razor-thin disk model of \cite{CMO2022} provides a good opportunity to test the collective motion of particles in the vicinity of the Kerr black hole. Novel elements of this model, covered in this work, include a discussion of monoenergetic flows and the properties of the bulk angular velocity, deviating from ZAMO values.

The Monte Carlo method proposed in \cite{MCO2023,CMO2024} and developed in this paper consists of the following three elements: (i) selecting a sample of geodesic parameters, reflecting the assumed asymptotic conditions, (ii) solving the geodesic equations, (iii) computing suitable averages to provide Monte Carlo estimators of observable quantities. In examples considered in this paper, the selection of geodesic parameters corresponding to planar models is relatively straightforward, mostly because the values of the azimuthal angular momentum can be selected from a uniform distribution. The assumed axial symmetry and a suitably defined averaging procedure render the computation of a majority of elements of individual particle trajectories unnecessary. In the examples shown in this paper, it is actually sufficient to compute the location of radial turning points for scattered orbits. For less symmetric flows, a computation of individual geodesic orbits would still be required. In general, stationary solutions can be recovered without solving for the time coordinate along individual trajectories. The employed averaging procedure consists in counting (with appropriate weights) intersections of individual particle orbits with surfaces adapted to the symmetry of the problem. In this work, we followed the choice of \cite{CMO2024} and considered intersections of orbits with circles of a constant coordinate radius.

The analysis of this paper neglects the electromagnetic fields, and hence it is rather applicable to models of dark matter particles (in the spirit of \cite{pmao1,mms2025a}) than to magnetized plasmas.  On the other hand, a very simple implementation of our Monte Carlo scheme in the case of the Kerr spacetime suggests that the method is robust and can be extended to more complex scenarios, including electromagnetic fields (replacing geodesic motion with the motion of test charges in a given electromagnetic field).

\begin{acknowledgments}
P.\ M.\ acknowledges a support of the Polish National Science Centre Grant No.\ 2017/26/A/ST2/00530.
\end{acknowledgments}

\end{document}